\documentclass[aps,prl,citeautoscript,twocolumn,reprint,longbibliography,floatfix,superscriptaddress]{revtex4-2}
\usepackage{graphicx,amssymb,amsmath,epsf,bm,cprotect,comment,physics, eqnarray}
\epsfclipon
\usepackage{comment,physics}

\newcommand{\unit}[1]{\,\mathrm{#1}}

\newcommand{\expct}[1]{\langle{#1}\rangle}

\newcommand{\prt}[2]{\frac{\partial{#1}}{\partial{#2}}}

\makeatletter
\renewcommand{\eqref}[1]{Eq.\,(\ref{#1})}

\newcommand{\pref}[1]{(\ref{#1})}
\newcommand{\figref}[2][\@empty]{Fig.\,\ref{#2}\ifx\@empty#1{}\else{(\uppercase{#1})}\fi}
\newcommand{\fref}[2][\@empty]{\ref{#2}\ifx\@empty#1{}\else{(\uppercase{#1})}\fi}

\newcommand{\supfigref}[2][\@empty]{Fig.\,\ref{#2}\ifx\@empty#1{}\else{(\uppercase{#1})}\fi}

\newcommand{\vidref}[1]{Movie\,#1}
\newcommand{\vidsref}[1]{Movies\,#1}
\makeatother

\usepackage{xr} %external reference
\usepackage{xcolor}
\makeatletter
\newcommand*{\addFileDependency}[1]{
  \typeout{(#1)}
  \@addtofilelist{#1}
  \IfFileExists{#1}{}{\typeout{No file #1.}}
}
\makeatother

\newcommand*{\myexternaldocument}[1]{
    \externaldocument[S-]{build/#1}  % for arXiv
    \addFileDependency{#1.tex}
    \addFileDependency{build/#1.aux}  % for arXiv
}
\myexternaldocument{suppl-arxiv}

\usepackage{ascmac}

\begin{document}

%\title{Emergence of bacterial glass: two-step glass transition in 2D bacterial suspension}
\title{Emergence of bacterial glass}

\author{Hisay Lama}
%\affiliation{Department of Physics,\! The University of Tokyo,\! 7-3-1 Hongo,\! Bunkyo-ku,\! Tokyo 113-0033,\! Japan}%
\affiliation{Department of Physics, The University of Tokyo, Tokyo, Japan}%

\author{Masahiro J. Yamamoto}
\affiliation{National Metrology Institute of Japan, National Institute of Advanced Industrial Science and Technology, Tsukuba, Japan}
\affiliation{Department of Physics, The University of Tokyo, Tokyo, Japan}%

\author{Yujiro Furuta}
%\affiliation{Department of Physics,\! Tokyo Metropolitan University,\! 1-1 Minami-Oosawa,\! Hachioji,\! Tokyo 192-0397,\! Japan}
\affiliation{Department of Physics, Tokyo Metropolitan University, Tokyo, Japan}%
%\affiliation{Department of Physics,\! Tokyo Institute of Technology,\! 2-12-1 Ookayama,\! Meguro-ku,\! Tokyo 152-8551,\! Japan}
\affiliation{Department of Physics, Tokyo Institute of Technology, Tokyo, Japan}%

\author{Takuro Shimaya}
\affiliation{Department of Physics, The University of Tokyo, Tokyo, Japan}%
\affiliation{Department of Physics, Tokyo Institute of Technology, Tokyo, Japan}%

\author{Kazumasa A. Takeuchi}
\email{kat@kaztake.org}
\affiliation{Department of Physics, The University of Tokyo, Tokyo, Japan}%
\affiliation{Department of Physics, Tokyo Institute of Technology, Tokyo, Japan}%
\date{\today}

%\textbf{
\begin{abstract}
Densely packed, motile bacteria can adopt collective states not seen in conventional, passive materials. These states remain in many ways mysterious, and their physical characterization can aid our understanding of natural bacterial colonies and biofilms as well as materials in general. 
Here, we overcome challenges associated with generating uniformly growing, large, quasi-two-dimensional bacterial assemblies
%Here we overcome this 
by a membrane-based microfluidic device and report the emergence of glassy states in two-dimensional suspension of \textit{Escherichia coli}. As the number density increases by cell growth, populations of motile bacteria transition to a glassy state, where cells are packed and unable to move. This takes place in two steps, the first one suppressing only the orientational modes and the second one vitrifying the motion completely. 
Characterizing each phase through statistical analyses and investigations of individual motion of bacteria, we find not only characteristic features of glass such as rapid slowdown, dynamic heterogeneity and cage effects, but also a few properties distinguished from those of thermal glass.
These distinctive properties include the spontaneous formation of micro-domains of aligned cells with collective motion, the appearance of an unusual signal in the dynamic susceptibility, and the dynamic slowdown with a density dependence generally forbidden for thermal systems. 
Our results are expected to capture general characteristics of such active rod glass, which may serve as a physical mechanism underlying dense bacterial aggregates.
\end{abstract}
\maketitle

\begin{itembox}[l]{Significance statement}
Bacteria often live in the form of dense populations, such as biofilms. 
While diverse approaches have been taken to understand such aggregates, physical consequences of being dense remained largely unexplored. 
Here, by using a microfluidic device suitable for a uniform culture of dense bacteria, we revealed that bacteria transition from an actively swimming state to jammed states as they proliferate, through a pathway analogous to glass transitions of colloidal rods. 
Through analysis of both single-cell and statistical properties, we characterized the observed collective states and transitions, and identified not only similarities but dissimilarities with usual glass formers including colloids. 
Our model experiment of dense bacteria may impact broad contexts beyond biofilms, hinting at general characteristics of such active rod systems.
% 120 words = upper bound
\end{itembox}

%%%%%%%%%%%%%%%%%%%%%%%%%
\begin{figure*}[t]
\centering
\includegraphics[width=\hsize]{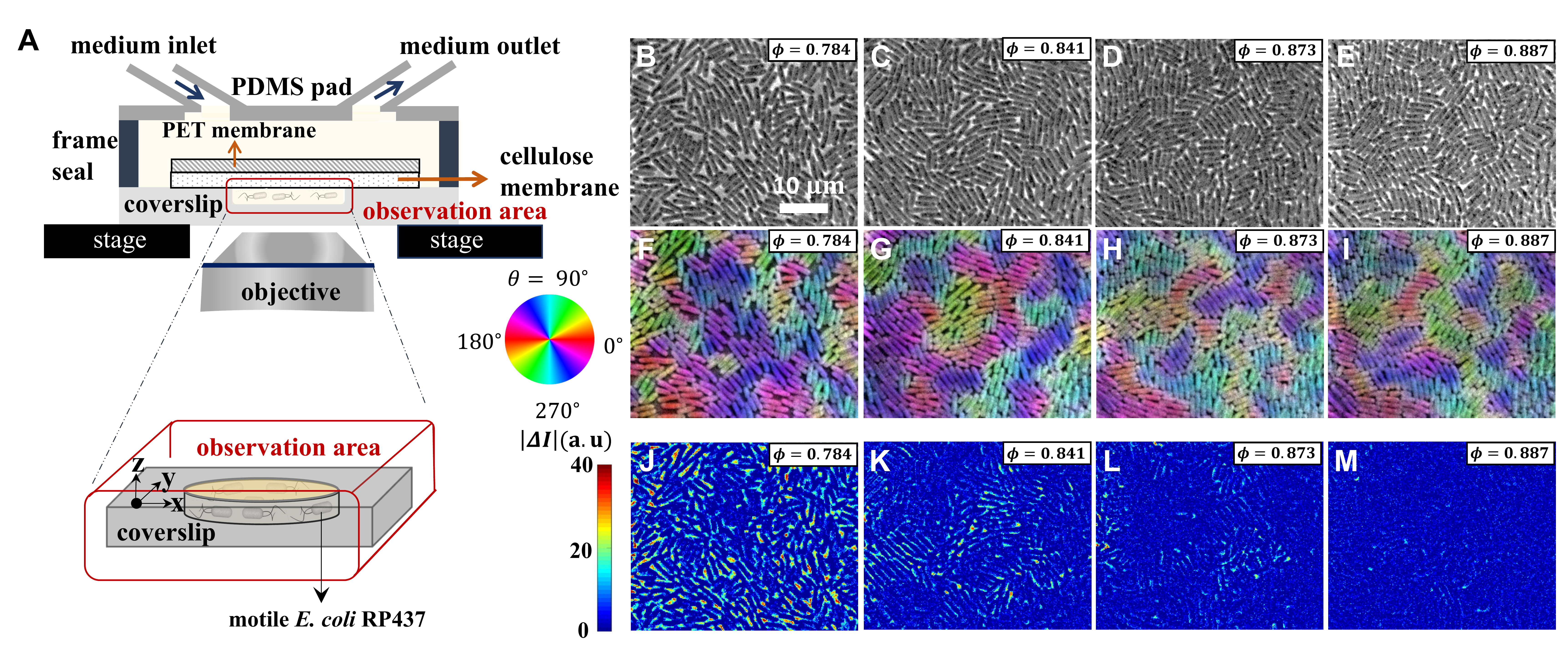}
\caption{
Experimental setup and vitrification of the bacterial population. A) Sketch of the experimental system.
B-E) Phase-contrast images for different area fractions $\phi$, taken at different times from a single experimental run (see \supfigref{S-fig:density}).
F-I) Orientation field (pseudocolor) overlaid on the phase-contrast images.
J-M Intensity difference $\Delta I(\vec{r}, t, \Delta t) = I(\vec{r}, t + \Delta t) - I(\vec{r},t)$ with $\Delta t = 0.053\unit{s}$.
}
\label{fig1}
\end{figure*}
%%%%%%%%%%%%%%%%%%%%%%%%

\section*{Introduction}

%\dropcap{B}acteria on Earth often live in the form of dense aggregates, i.e., biofilms \cite{Flemming.Wuertz-NRM2019,Sauer.etal-NRM2022}, which are also encountered in medicine and industries and are often the bane \cite{HallStoodley.etal-NRM2004,Vishwakarma-JBM2019,Sauer.etal-NRM2022}.
%Whereas understanding biofilms certainly requires characterizing their multifaceted aspects \cite{Flemming.Wuertz-NRM2019,Sauer.etal-NRM2022}, including extracellular substances and cell phenotypes, physical consequences of being dense aggregates receive less attention.
Dense bacterial populations offer an exciting frontier of research for both physical and microbial sciences. 
From the biological viewpoint, dense bacterial populations, especially biofilms, abound in diverse natural environments and beyond \cite{Flemming.Wuertz-NRM2019,Sauer.etal-NRM2022}, being also encountered in medicine and industries often as the bane \cite{HallStoodley.etal-NRM2004,Vishwakarma-JBM2019,Sauer.etal-NRM2022}. 
It is therefore crucial to characterize their multifaceted aspects \cite{Flemming.Wuertz-NRM2019,Sauer.etal-NRM2022}, including not only extracellular substances and cell phenotypes as studied extensively in the literature, but also physical consequences of being dense aggregates, which have gained an emerging interest in physical sciences. 
Indeed, recent physical studies on dense bacterial populations unveiled a plethora of collective states not seen in conventional passive materials \cite{Beer.etal-CP2020, Aranson-RPP2022}, which however remain in many ways mysterious.

Generally, dense particle systems may undergo glass and jamming transitions as the number density is increased, showing dramatic changes in both single-particle dynamics and material properties as suspension \cite{Gotze-book,berthier2011theoretical,hunter2012physics, Charbonneau.etal-ARCMP2017, Reichman.Charbonneau-JSM2005}.
Even for conventional thermal systems, it is only recently that firm theoretical grounds on glass transitions started to be built for some idealized cases \cite{Parisi.etal-book2019}. 
Therefore, it constitutes an important challenge in physics to extend this understanding to active systems \cite{janssen2019active,Berthier.etal-JCP2019,Sadhukhan.etal-a2024}, i.e., systems made of motile particles akin to cells, or more generally, to athermal systems including intracellular environments. 
%This phenomenon is even richer in active systems \cite{janssen2019active,Berthier.etal-JCP2019}, for which self-propulsion of constituent particles adds a nontrivial effect on glassy dynamics.
%Characterization of glassy dynamics and transitions in dense bacterial populations has been faced with challenges.
%First, unlike thermal systems made of, e.g., colloidal particles, bacterial cells grow and self-propel.
%Such active characters of particles are known to have nontrivial effects upon glass transitions, leading to the recent surge of interest in active glass systems \cite{janssen2019active,Berthier.etal-JCP2019}. 
Experimentally, glassy dynamics has been reported in cytoplasm and cell extracts \cite{Zhou.etal-PNAS2009,Parry.etal-C2014,Nishizawa.etal-SR2017} as well as in mammalian tissues \cite{angelini2011glass,park2015unjamming,Garcia.etal-PNAS2015}, posing interesting problems both in physics and biology \cite{janssen2019active,Berthier.etal-JCP2019,Sadhukhan.etal-a2024}. 
Concerning bacteria, by contrast, whereas a few recent studies suggested the relevance of glassy phenomena to the regulation of motility and three-dimensional growth in bacterial colonies \cite{Rhodeland.etal-JRSI2020,takatori2020motility}, identification of glassy states in bacterial populations remains a challenge, especially under controlled environments that are necessary for the quantitative investigation. 
A primary difficulty is to keep uniform growth conditions for such dense cell populations for a long time. 
While microfluidics is generally suitable for controlled experiments, conventional devices that deliver nutrients through channels cannot maintain uniform growth conditions for large and dense populations \cite{Mather.etal-PRL2010}.

Here we overcame past difficulties and report a controlled experiment characterizing emergent glassy phases of dense populations of motile bacteria. 
We used a membrane-type microfluidic device developed in \cite{Shimaya2021} (\figref[a]{fig1}), named the extensive microperfusion system (EMPS). 
Delivering growth medium to bacteria through a porous membrane, this device can maintain a uniform and constant growth condition for bacteria trapped in a 2D well, even if the bacterial cells are densely packed \cite{Shimaya2021}.
The membrane may also let out substances secreted by cells, thus providing an ideal platform for studying the physical effect of crowding.
In the present work, we cultured motile bacteria (\textit{E. coli}, strain RP437, width $\approx 1\unit{\mu m}$, length varying roughly from $2$ to $6\unit{\mu m}$; see \supfigref{S-fig:polydispersity} for the distribution of cell areas) in a closed 2D well (diameter $71.2 \pm 0.5 \unit{\mu m}$, depth $\approx 1.4\unit{\mu m}$) supplied with growth medium (tryptone broth supplemented with surfactant; see Methods), and monitored their spatiotemporal dynamics in a region near the center. 
Bacteria were initially swimming actively. 
However, as the area fraction of bacteria, $\phi$, increased by cell growth and division at a uniform and constant growth rate (\figref[b-e]{fig1}, \supfigref{S-fig:density} and \supfigref{S-fig:SpatialPhi}), we found that the motion of the bacterial population started to be hampered rapidly (\vidsref{1-5}), while their positions and orientations remained globally disordered in space (\figref[f-i]{fig1}).
In particular, the static structure factor does not show a significant change in its shape (\figref[a]{fig2}) in the range of $\phi$ where the bacterial motion slows down. 
%We then find that the bacteria eventually vitrify from an actively swimming state to an amorphous, jammed state, showing rapid dynamical slowdown and other hallmarks of glass transitions including the dynamic heterogeneity and cage escape events. 
%Notably, the glass transition turns out to take place in two steps, the first one suppressing only the orientational degrees of freedom of bacteria, and the second one vitrifying the bacteria completely. 
%We also show that bacteria spontaneously form micro-domains of cells aligned with each other, leading to collective motion which seems to add a characteristic signal to statistical properties of the glass transition. 
%Note that, as a particulate system, our \textit{E. coli} population can be regarded as a polydisperse mixture of spherocylinders of width $\approx 1\unit{\mu m}$ and length varying from $\approx 2\unit{\mu m}$ to $\approx 6\unit{\mu m}$ (see \supfigref{S-fig:polydispersity} for the distribution of cell areas, with mean $4.08\unit{\mu m^2}$ and polydispersity index $1.07$), which self-propel by rotating their flagella.

\section*{Rapid dynamic slowdown} 

The rapid suppression of motion can be evaluated by the difference of phase-contrast image intensities taken at two different times, $\Delta I(\vec{r}, t, \Delta t) = I(\vec{r}, t + \Delta t) - I(\vec{r},t)$. Figure\,\fref[j-m]{fig1} shows that, with increasing $\phi$, the region with $\Delta I(\vec{r}, t, \Delta t) \approx 0$ (where bacteria hardly moved during the chosen time interval) expanded, and eventually, near $\phi \approx 0.88$, the entire population became kinetically arrested, i.e., vitrified. 
We also notice that this kinetic arrest took place heterogeneously (see \figref[k,l]{fig1} and \vidref{1}), analogously to glassy systems showing the dynamic heterogeneity \cite{berthier2011theoretical,hunter2012physics}.

%%new-content LAMA
%{\color{blue} We characterize the observed vitrification by evaluating the static structure factor, $S(q)$, which is computed as the modulus squared of the space Fourier transform of $I(\Vec{r},t)$, here, $q$ represents the wave vector. Figure\,\fref[a]{fig2} shows the decay of \textit{S(q)} in the reciprocal space within the investigated range of $\phi$, indicating the absence of long-range structural correlations. The observation confirms the lack of spatial order, which is further supported by the time-averaged Fourier transform images (see inset of \figref[a]{fig2}). Despite the apparent disordered phase in dense bacterial populations, the structural similarity and isotropic characteristics of a system's dilute and dense phases emphasize the need to investigate its dynamics for understanding vitrification.}
%Therefore, we characterize the glassy dynamics
%%%%%%%%%%%%%%%%%%%%%%%%%
\begin{figure*}[t]
    \centering
    \includegraphics[width=\linewidth]{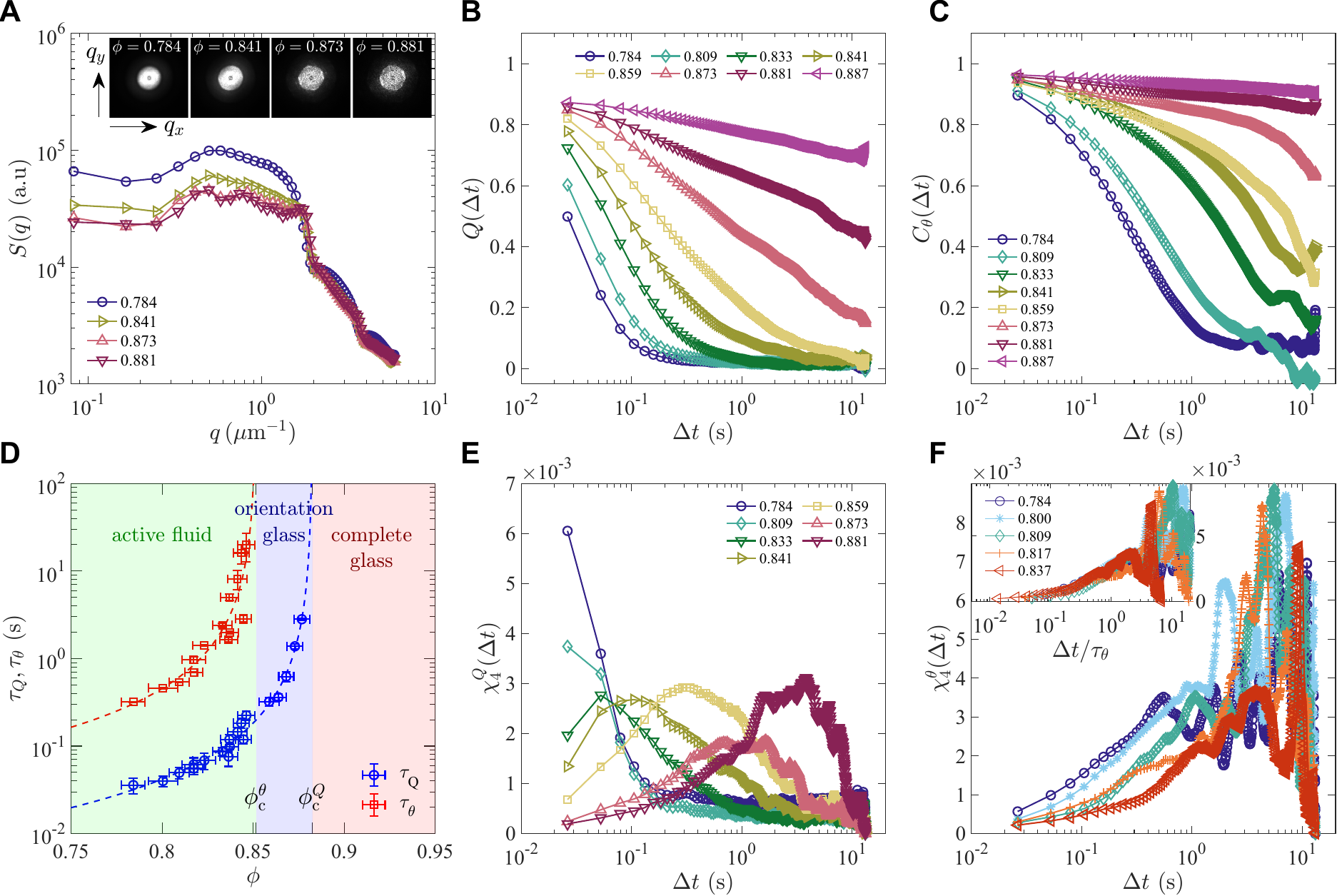}
    \caption{
    Static structure factor, structural relaxation, and dynamic susceptibility. A) The static structure factor $S(\vec{q})$ for different area fractions $\phi$. In the main panel, it is shown as a function of $q = |\vec{q}|$ by taking the average over all angles in the reciprocal space.
    B) The overlap function $Q(\Delta t)$ for different area fractions $\phi$ (legend).
    C) The orientational correlation function $C_\theta(\Delta t)$ for different area fractions $\phi$ (legend).
    D) The translational (overlap) and the orientational relaxation times, $\tau_Q$ (blue) and $\tau_\theta$ (red), respectively, as functions of $\phi$. The dashed lines indicate the results of the MCT fitting, $\tau \sim (\phi_\mathrm{c} - \phi)^{-\gamma}$, with $\phi_\mathrm{c}^Q = 0.882(4)$ for $\tau_Q$ and $\phi_\mathrm{c}^\theta = 0.852(13)$ for $\tau_\theta$.
    E) The dynamic susceptibility $\chi_4^Q(\Delta t)$ associated with the overlap function, for different area fractions $\phi$ (legend).
    F) The dynamic susceptibility $\chi_4^\theta(\Delta t)$ associated with the orientation modes, for different area fractions $\phi$ (legend). The same symbols indicate the same values of $\phi$ in all panels except D.
    }
    % \textbf{Structural relaxation and dynamic susceptibility. a,} The overlap function $Q(\Delta t)$ for different area fractions $\phi$ (legend).
    % \textbf{b,} The orientational correlation function $C_\theta(\Delta t)$ for different area fractions $\phi$ (legend).
    % \textbf{c,} The translational (overlap) and the orientational relaxation times, $\tau_Q$ (blue) and $\tau_\theta$ (red), respectively, as functions of $\phi$. The dashed lines indicate the results of the MCT fitting, $\tau \sim (\phi_\mathrm{c} - \phi)^{-\gamma}$, with $\phi_\mathrm{c}^Q = 0.882(4)$ for $\tau_Q$ and $\phi_\mathrm{c}^\theta = 0.851(11)$ for $\tau_\theta$.
    % \textbf{d,} The dynamic susceptibility $\chi_4(\Delta t)$ for different area fractions $\phi$ (legend). The same symbols are used in panels \textbf{a,b,d}.
    % }
    \label{fig2}
\end{figure*}
%%%%%%%%%%%%%%%%%%%%%%%%%%%%

We characterize the observed vitrification by the differential variance analysis (DVA) \cite{pastore2017differential}, 
%We characterize the relaxation dynamics of a vitrifying system by the differential variance analysis (DVA) \cite{pastore2017differential}, 
which uses the intensity difference $\Delta I(\vec{r}, t, \Delta t)$ to analyze the structural relaxation and the dynamic heterogeneity.
The structural relaxation is studied by the overlap function defined by
\begin{equation}
    Q(\Delta t) = 1 - \frac{V(\Delta t)}{V(\infty)},  \label{eq:DefQ}
\end{equation}
where $V(\Delta t) = \langle {\Delta I(\vec{r}, t, \Delta t)^2}\rangle_{\vec{r},t}$ is the intensity variance taken over position $\vec{r}$ and reference time $t$, and $V(\infty)$ is evaluated by twice the variance of $I(\vec{r},t)$. 
The quantity $Q(\Delta t)$ roughly corresponds to the fraction of bacteria that did not move over lag time $\Delta t$.
More quantitatively, $Q(\Delta t)$ was reported to behave similarly to the self-intermediate scattering function \cite{pastore2017differential}, a quantity often used to characterize the structural relaxation of glassy materials.

Figure\,\fref[b]{fig2} shows the result of structural relaxation assessed through $Q(\Delta t)$ for different $\phi$.
For low $\phi$, $Q(\Delta t)$ decays to zero after a relatively short relaxation time, indicating fast structural relaxation.
This corresponds to what is called the $\alpha$-relaxation in the literature \cite{Gotze-book,berthier2011theoretical,hunter2012physics, Charbonneau.etal-ARCMP2017, Reichman.Charbonneau-JSM2005}.
However, this $\alpha$-relaxation time increases rapidly for $\phi \gtrsim 0.85$, soon exceeding the observation time.
To be more quantitative, we fit the data by a stretched exponential function, $Q(\Delta t) \sim e^{-(\Delta t/\tau_Q)^{\beta_Q}}$, well-known to describe the $\alpha$-relaxation of glassy materials \cite{Gotze-book,berthier2011theoretical}, and evaluate the relaxation time $\tau_Q$ thereby
%(\efigref{fig:StretchedExpQ}).
(\supfigref{S-fig:StretchedExpQ}).
The result indeed shows a rapid increase of $\tau_Q$ (\figref[d]{fig2} blue symbols), by nearly two orders of magnitude in $0.78 \lesssim \phi \lesssim 0.88$.
The observed superexponential growth of $\tau_Q$ indicates that our bacterial populations are a fragile glass former \cite{Gotze-book,berthier2011theoretical,hunter2012physics}.
It is compatible with typical growth laws documented in the literature, such as the power-law divergence 
%$\tau_Q \sim (\phi_\mathrm{c}^Q - \phi)^{-\gamma_Q}$ 
\begin{equation}
    \tau_Q \sim (\phi_\mathrm{c}^Q - \phi)^{-\gamma_Q}  \label{eq:tau_Q}
\end{equation}
predicted by mode-coupling theories (MCT) \cite{Gotze-book,berthier2011theoretical,Reichman.Charbonneau-JSM2005} (\figref[d]{fig2} blue dashed line), as well as the Vogel-Fulcher-Tamman law \cite{Gotze-book,berthier2011theoretical,hunter2012physics}, $\tau \sim \exp\left(\frac{c\phi}{\phi_\mathrm{VFT}-\phi}\right)$ (\supfigref{S-fig:VFT}).
This allows us to evaluate the glass transition point; for example, from the MCT power law \pref{eq:tau_Q}, one obtains $\phi_\mathrm{c}^Q = 0.882(4)$, where the number(s) in the parentheses represents the uncertainty in the last digit(s) (see Methods). 

%%%%%%%%%%%%%%%%%%%%%%%%%%
\begin{figure*}[t]
    \centering
    \includegraphics[width=\hsize]{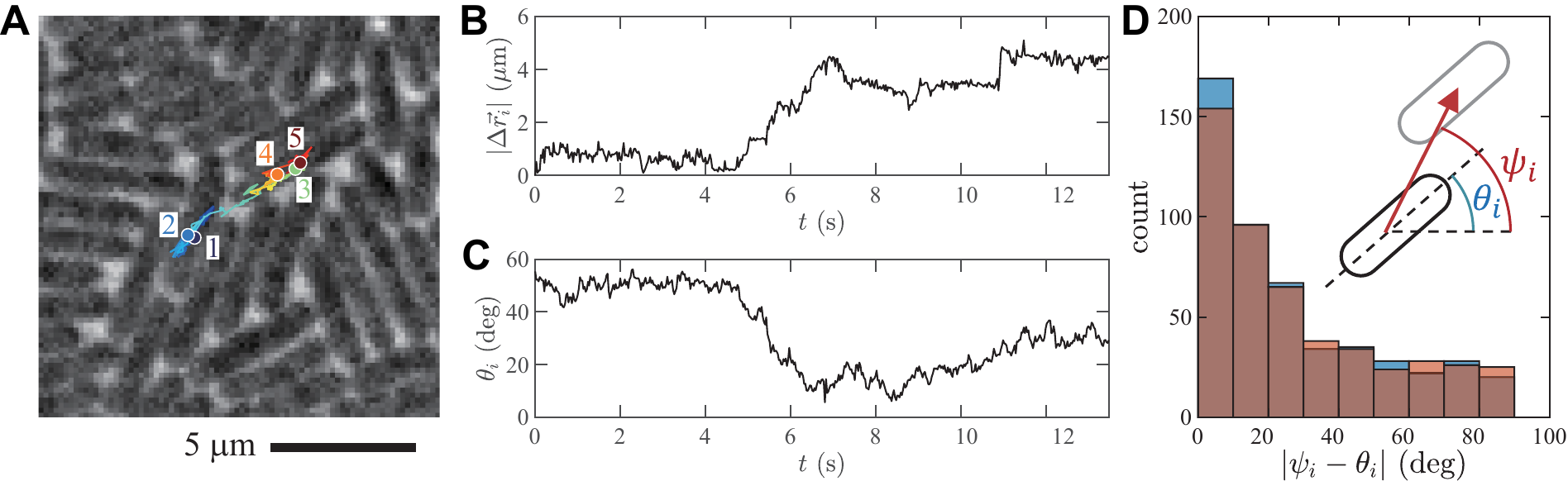}
    \caption{
    Glassy dynamics in the motion of individual bacteria.
    The displayed results are for $\phi = 0.873(4)$, which is in the orientation glass phase.
    A) Trajectory of a single cell for $0 \leq t \leq 13.14\unit{s}$, drawn on the phase-contrast image taken at the last time frame. The positions at $t = 0, 3.29, 6.58, 9.87, 13.14\unit{s}$ are shown by colored disks with labels $1, 2, \cdots, 5$, respectively.
    B,C) Time series of the displacement from the initial position, $|\Delta\vec{r}_i(t)|$ (B), and that of the orientation $\theta_i(t)$ (C) of the cell tracked in panel A.
    These time series show a cage escape event during $5\unit{s} \lesssim t \lesssim 7\unit{s}$.
    D) Histogram of the angle difference between the cell orientation and the displacement (see the sketch). The cell orientations and the displacements are measured between two times separated by an interval $\Delta t$, with $\Delta t=0.026\unit{s}$ (blue) and $0.26\unit{s}$ (red).
    The two histograms are overlaid with semitransparent colors, so that the dark bars indicate the overlapping part and the light blue and red regions indicate the difference.
    }
    \label{fig3}
\end{figure*}
%%%%%%%%%%%%%%%%%%%%%%%%%%%%

\section*{Two-step transition} 

At this point, it is worth recalling the spherocylindrical shape of the constituting entity, namely \textit{E. coli}, which has both translational and orientational degrees of freedom.
The relaxation of the orientational degrees of freedom can be evaluated by the orientational correlation function
\begin{equation}
%    C_\theta (\Delta t) = \langle \vec{n}_{2\theta}(\theta(\vec{r}, t + \Delta t) . \vec{n}_{2\theta}(\theta(\vec{r}, t) \rangle_{\vec{r},t}
    C_\theta (\Delta t) = \langle \cos 2[\theta(\vec{r}, t + \Delta t) - \theta(\vec{r}, t)] \rangle_{\vec{r},t},  \label{eq:DefCtheta}
\end{equation}
where $\theta(\vec{r},t)$ represents the nematic orientation angle. 
Figure\,\fref[c]{fig2} shows $C_\theta(\Delta t)$ for different $\phi$. 
Similarly to $Q(\Delta t)$, $C_\theta(\Delta t)$ also decays, following the stretched exponential form $C_\theta(\Delta t) \sim e^{-(\Delta t/\tau_\theta)^{\beta_\theta}}$ (\supfigref{S-fig:StretchedExpTheta}), with a characteristic relaxation time $\tau_\theta$ that increases rapidly with $\phi$ (\figref[d]{fig2} red symbols).
Importantly, we find that the orientational relaxation time $\tau_\theta$ is larger than that of the overlap function $\tau_Q$ by an order of magnitude or more, and seems to diverge at lower $\phi$.
%This is underpinned by the MCT fitting $\tau_\theta \sim (\phi_\mathrm{c}^\theta - \phi)^{-\gamma_\theta}$ (\figref[d]{fig2} red dashed line), which gives $\phi_\mathrm{c}^\theta = 0.851(11)$ that is significantly smaller than $\phi_\mathrm{c}^Q = 0.882(4)$.
This is underpinned by the MCT fitting (\figref[d]{fig2} red dashed line)
\begin{equation}
    \tau_\theta \sim (\phi_\mathrm{c}^\theta - \phi)^{-\gamma_\theta},  \label{eq:tau_theta}
\end{equation}
which gives $\phi_\mathrm{c}^\theta = 0.852(13)$ that is significantly smaller than $\phi_\mathrm{c}^Q = 0.882(4)$.
The same conclusion was reached when the data were fitted with the Vogel-Fulcher-Tamman law (\supfigref{S-fig:VFT}).
From $\phi_\mathrm{c}^\theta < \phi_\mathrm{c}^Q$, we conclude that the orientational degrees of freedom vitrify earlier than the rest, i.e., the translational degrees of freedom, the latter of which essentially governed the relaxation of the overlap function.
In other words, the glass transition in our system takes place in two steps, the first being a transition to the \textit{orientation glass} at $\phi_\mathrm{c}^\theta = 0.852(13)$ and the second the ultimate transition to the \textit{complete glass} at $\phi_\mathrm{c}^Q = 0.882(4)$ (\figref[d]{fig2}).
%A similar two-step transition was also reported for glass transitions of ellipsoidal colloids \cite{zheng2011glass,Mishra.etal-PRL2013, zheng2014structural,Roller.etal-PNAS2021}, while there also exist theoretical and numerical studies which showed that translational degrees of freedom may vitrify before the orientational ones \cite{Letz.etal-PRE2000,Mandal.etal-PRE2017}.
%It is also interesting to note that our estimates of the exponent $\gamma$, specifically $\gamma_Q = 1.5(3)$ and $\gamma_\theta=1.5(12)$, seem to suggest the athermal nature of our system, because MCT generally gives $\gamma > 1.76\dots$ for thermal systems \cite{Gotze-book} (see Appendix B).

\section*{Dynamic heterogeneity} 

Another hallmark of glassy dynamics is dynamic heterogeneity, which concerns nontrivial spatiotemporal correlation that develops near the glass transition \cite{berthier2011theoretical,hunter2012physics}.
It is often quantified by the dynamic susceptibility $\chi_4$, which is essentially the variance of the structural correlation function.
In DVA \cite{pastore2017differential}, it can be evaluated through the variance of $Q(t, \Delta t)$, defined analogously to \eqref{eq:DefQ} but with a given $t$, without taking average over time.
Here we adopt the following definition:
\begin{equation}
    \chi_4^Q (\Delta t) = \phi [\langle Q(t, \Delta t)^2\rangle_t - \langle Q(t, \Delta t) \rangle_t^2],  \label{eq:DefChi4}
\end{equation}
where $\langle\cdots\rangle_t$ denotes time averaging.
The result in \figref[e]{fig2} shows that $\chi_4^Q(\Delta t)$ develops a peak as the second transition point $\phi_\mathrm{c}^Q = 0.882(4)$ is approached, at the time scale consistent with $\tau_Q$.
This is typical of glassy systems \cite{berthier2011theoretical}, providing another support of characteristic glassy dynamics in our bacteria.
By contrast, \figref[e]{fig2} also shows an unusual peak development for low $\phi$, at small $\Delta t$.
This will be scrutinized below, through the analysis of the motion of bacteria and the collectivity.
We also evaluated the dynamic susceptibility associated with the orientational degrees of freedom, defined by $\chi_4^\theta (\Delta t)$ = $\phi [\langle C_\theta(t, \Delta t)^2\rangle - \langle C_\theta(t, \Delta t)\rangle^2]$ (\figref[f]{fig2}).
In this quantity, we only see the peak due to the transition to the orientation glass, with the profile change controlled by the increase of $\tau_\theta$ (inset).

%%%%%%%%%%%%%%%%%%%%%%%%%%
\begin{figure*}[t]
    \centering
    \includegraphics[width=\hsize]{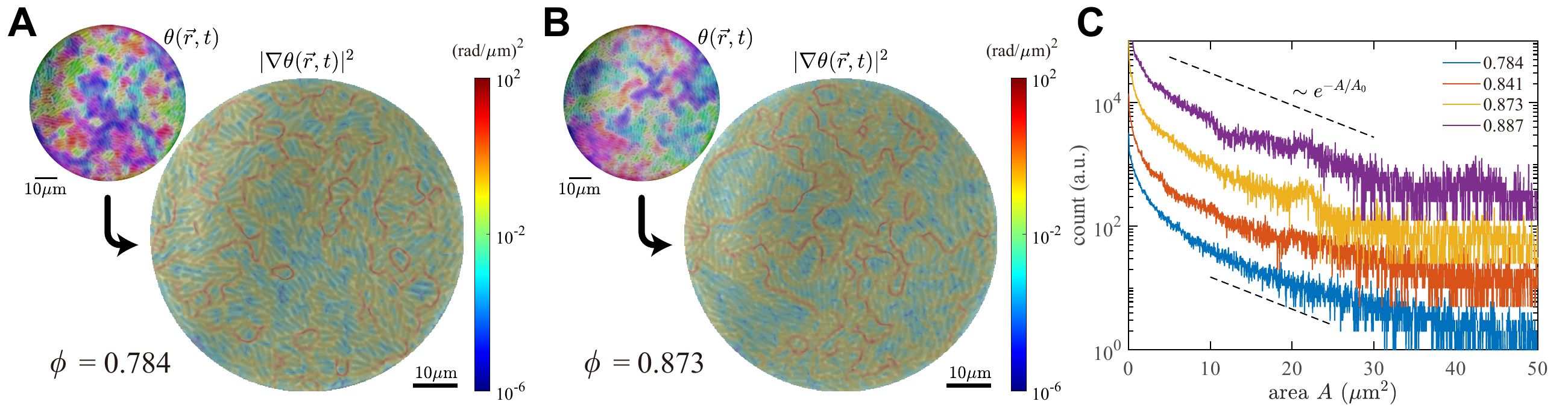}
    \caption{
    Formation of microdomains of aligned cells. A,B) The orientation field $\theta(\vec{r},t)$ (top left) and its gradient squared $|\nabla\theta(\vec{r},t)|^2$ (main) overlaid on the phase-contrast image, for $\phi = 0.784(7)$ (A, active fluid) and $\phi = 0.873(4)$ (B, orientation glass). 
    See also \vidsref{8-11}.
    The pseudocolor code for the orientation is identical to that used in \figref[F-I]{fig1}.
    C) Histograms of the microdomain area $A$ for different $\phi$ (legend). The results were shifted vertically for the sake of visibility. The dashed lines are guides for the eyes indicating the exponential distribution $e^{-A/A_0}$ with $A_0 = 8.3(21)\unit{\mu m^2}$. The presented results were obtained with a threshold $|\nabla\theta(\vec{r},t)| \geq 10^{-2}\unit{(rad/\mu m)^2}$ here. Changing this threshold results in a constant vertical shift (\supfigref{S-fig:DomainAreaDist}), while it hardly affects the characteristic area $A_0$ of the exponential distribution.
    }
    \label{fig4}
\end{figure*}
%%%%%%%%%%%%%%%%%%%%%%%%%%%%

\section*{Individual dynamics of bacteria} 

Now we characterize the observed phases.
First, we tracked single cells and investigated the evolution of their position $\vec{r}_i(t)$ and orientation $\theta_i(t)$.
Figure\,\fref[a]{fig3} displays an example in the orientation glass phase, shown with time series of the displacement $\Delta\vec{r}_i(t) = \vec{r}_i(t)-\vec{r}_i(0)$ and the orientation $\theta_i(t)$ (\figref[b,c]{fig3}, respectively; see also \vidref{6}).
This cell was initially caged by neighbors ($t \lesssim 5\unit{s}$), but eventually escaped and moved significantly, over $4\unit{\mu m}$ or so during $5\unit{s} \lesssim t \lesssim 7\unit{s}$, until it was caged again.
This is the cage effect, another characteristic of glassy systems \cite{berthier2011theoretical,hunter2012physics,Gotze-book} (see \supfigref{S-fig:cage2} and \vidref{7} for another cage escape event).
Moreover, \figref[d]{fig3} shows that the displacement tends to occur along the cell orientation, even in such a dense population where cells are pushed by neighbors in all directions.
In other words, cells are led to move along the orientation field $\theta(\vec{r},t)$, which is frozen in the orientation glass.
As a result, it is practically only when bacteria move, by escaping from a cage, that they change the orientation significantly in the orientation glass phase (compare \figref[bc]{fig3}). 
By contrast, in the active fluid phase $\phi < \phi_\mathrm{c}^\theta$, the orientation field $\theta(\vec{r},t)$ along which bacteria tend to move evolves in space and time.

%%%%%%%%%%%%%%%%%%%%%%%%%%
\begin{figure*}[t]
    \centering
    \includegraphics[width=\hsize]{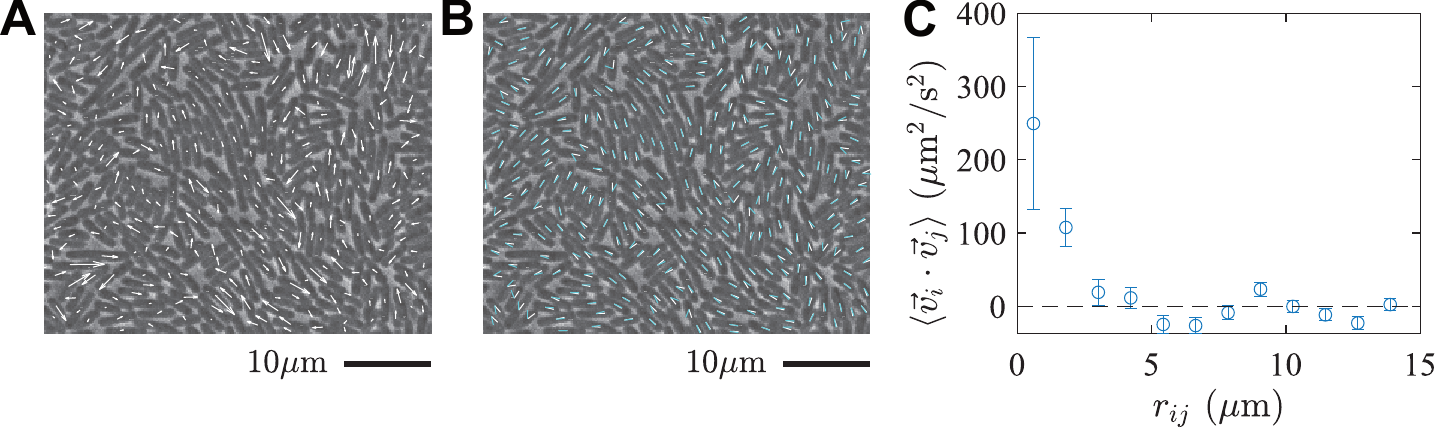}
    \caption{
    Collective motion of bacteria in the active fluid phase for $\phi = 0.784(7)$. A,B) The displacements $\vec{r}_i(t+\Delta t) - \vec{r}_i(t)$ (A) and the orientation changes $\theta_i(t+\Delta t) - \theta_i(t)$ (B) of the individual bacteria during a short time interval, $\Delta t = 0.053\unit{s}$, displayed on top of the image at time $t$. In A, the displacements are indicated by the white arrows, whose endpoints are placed at the positions of the center of mass at times $t$ and $t+\Delta t$. In B, the orientation at time $t$ and $t+\Delta t$ is shown by the white and cyan lines, respectively. C) The velocity correlation $\expct{\vec{v}_i \cdot \vec{v}_j}$ as a function of distance $r_{ij} = |\vec{r}_j - \vec{r}_i|$.
    The distances $r_{ij}$ between cells were divided into bins and the average was taken in each bin.
    The error bars indicate the standard error, evaluated by regarding all pairs of cells as independent.
    }
    \label{fig5}
\end{figure*}
%%%%%%%%%%%%%%%%%%%%%%%%%%

\section*{Formation of microdomains and collective motion} 

Here we scrutinize the spatial structure of the orientation field $\theta(\vec{r},t)$, which turned out to guide the motion of bacteria.
Figures\,\fref[a,b]{fig4} show, in their top left corner, $\theta(\vec{r},t)$ measured in the entire well, for $\phi = 0.784(7)$ (active fluid phase) and $\phi = 0.873(4)$ (orientation glass), respectively (see also \vidsref{8-11}).
Both figures indicate the formation of nematic microdomains, composed of cells that are oriented in similar directions, reminiscent of those reported earlier for growing colonies of nonmotile bacteria \cite{You.etal-PRX2018}.
By plotting $|\nabla\theta(\vec{r},t)|^2$ (main panels of \figref[a,b]{fig4}), we find that these domains are surrounded largely, if not entirely, by nearly discontinuous borders (red; notice the logarithmic scale of the pseudocolor code) with neighboring domains oriented in different directions.
This microdomain structure is considered to be a result of the competition between the steric interaction, which tends to align the cells locally, and the activity-driven force, which originates from self-propulsion and cell growth and tends to destabilize the ordered alignment.
We characterize the distribution of microdomain areas $A$ and find an exponential distribution $e^{-A/A_0}$, with characteristic area $A_0 = 8.3(21)\unit{\mu m^2}$ that hardly depends on $\phi$ (\figref[c]{fig4} and \supfigref{S-fig:DomainAreaDist}).
The exponential distribution was also found for nonmotile bacterial colonies \cite{You.etal-PRX2018}, but the characteristic area $A_0$ turned out to be much smaller in our motile bacteria, presumably because of the stronger destabilizing force due to the self-propulsion.

An important consequence of the formation of microdomains is the emergence of collective motion.
Since individual bacteria tend to move along the orientation field, they attempt to move collectively in each microdomain, though the motion is largely hampered in the glassy states.
However, in the active fluid phase, bacteria do move in the form of microdomains or clusters (\figref[a,b]{fig5} and \vidsref{8 and 9}), more visibly for lower $\phi$, similarly to swarming states of bacteria \cite{marchetti2013hydrodynamics, bechinger2016active,doostmohammadi2019coherent, be2019statistical,be2020phase}.
The collective motion is evidenced by the presence of the velocity correlation between nearby cells (\figref[c]{fig5}), within a distance of $\approx 2\unit{\mu m}$, which is consistent with the typical size of the microdomains reported above.
Note that the velocity correlation is generally absent in thermal systems, hence another athermal characteristic of our system, and is known to play a significant role in active glassy dynamics \cite{janssen2019active,Szamel.etal-PRE2015,Sadhukhan.etal-a2024}.
In our system, the velocity correlation is formed as a result of the microdomain structure of the orientations, but it is noteworthy that the observed velocity correlation indicates the presence of polar order, despite the essentially nematic nature of the steric interaction. 
This collective motion produces nontrivial spatiotemporal correlation in the cell positions, which presumably results in the anomalous peak of the dynamic susceptibility $\chi_4^Q$ found to develop for low $\phi$ (\figref[e]{fig2}).
This argument is also supported by the fact that the anomalous peak is absent in the dynamic susceptibility of the orientation (\figref[f]{fig2}), $\chi_4^\theta$, which is not affected by the bacterial motion.
Interestingly, a similar short-time peak was also reported in a recent simulation of particles self-propelling in discrete directions and attributed to the cooperative vibrational motion of active particles \cite{Dey.etal-SM2022}.

\section*{Concluding remarks} 

To summarize, we report that dense populations of motile bacteria spontaneously vitrify as the number density increases by cell growth and division.
We found that this transition takes place in two steps as sketched in \figref[d]{fig2}, first from the active fluid phase to the orientation glass, where only the orientational degrees of freedom are arrested, and second to the complete glass where the remaining translational degrees of freedom are also arrested.
These transitions were found to show characteristic properties of glass transitions, such as rapid slowdown, fragility, dynamic heterogeneity, and cage effects. 
We also showed that bacteria form nematic microdomains, leading to collective motion which seems to result in the unusual signal in the dynamic susceptibility $\chi_4^Q$ for low $\phi$. 
These are attributed to the rod shape and self-propulsion of bacteria, and as such, can be characteristic of active rod systems in general.

Our analysis has revealed some other aspects that deserve further investigation. 
First, regarding the two-step transition, it is not trivial whether the glass transition should first occur in the orientation modes, in the translation modes, or both simultaneously. 
The literature is mixed in this respect; experiments and simulations of ellipsoidal colloid glass showed the two-step transition in the same order \cite{zheng2011glass,Mishra.etal-PRL2013, zheng2014structural,Roller.etal-PNAS2021}, but an MCT approach to hard ellipsoids \cite{Letz.etal-PRE2000} as well as simulations of an active dumbbell model \cite{Mandal.etal-PRE2017} exhibited the two transitions in the opposite order, i.e., first translation, then orientation.
The two transitions may occur simultaneously depending on the particle aspect ratio, but to our knowledge, no theory or system has demonstrated a change in the order.
It is therefore an interesting open problem to clarify what physical property controls the type of the two-step transition.
Second, we remark that our estimates of the MCT exponents for the dynamic slowdown, $\gamma_Q = 1.6(3)$ and $\gamma_\theta=1.5(13)$, 
%It is also interesting to note that our estimates of the exponent $\gamma$, specifically $\gamma_Q = 1.5(3)$ and $\gamma_\theta=1.5(12)$, 
seem to suggest the athermal nature of our system, because MCT generally gives $\gamma > 1.76\dots$ for thermal systems \cite{Gotze-book} (see Appendix B).
It is important to identify what aspect of our active system is responsible for violating this lower bound for thermal systems, and what the consequence is.

All in all, our experiment serves as a model system for investigating physical properties of dense bacterial populations, with potential relevance in various contexts such as biofilms. 
It also contributes to the understanding of characteristic glassy dynamics of such active rod systems, a class of active or athermal glass, which represents a frontier in the study of the physics of glass.

\section{Acknowledgments}
We are grateful to Y. Wakamoto and R. Okura for their help in the construction of a prototype of the experimental device used in this work. 
We thank A. Ikeda for his enlightening discussions on interpretations of the MCT fitting results (Appendix B), H. Yoshino and A. Ikeda for discussions on glass transitions of anisotropic particles, G. Luca for his suggestion to carry out domain analysis (\figref{fig4}), Y. Han for his suggestion to measure the displacements (\figref{fig5}), and D. Nishiguchi for frequent discussions. We acknowledge the useful codes and libraries made available by S. Panigrahi and L. Espinosa for MiSiC \cite{Panigrahi.etal-e2021}, by J. C. Crocker, E. R. Weeks, D. Blair and E. Dufresne for the particle tracking, by G. Xiong for the 3-class fuzzy c-means clustering \cite{FuzzyBook}, and by F. Grussu for the structure tensor method.

\section{Supplementary Material}
Supplementary material is available at PNAS Nexus online.

\section{Funding}
This work is supported in part by KAKENHI from Japan Society for the Promotion of Science (Grant Nos. JP16H04033, JP19H05800, JP20H00128, JP21K20350, JP24K00593), by Core-to-Core Program “Advanced core-to-core network for the physics of self-organizing active matter (JPJSCCA20230002), and by ``Planting Seeds for Research'' program and Suematsu Award from Tokyo Institute of Technology.

\section{Author contributions statement}
K.A.T. conceived the project and directed the research. H.L. and T.S. constructed the device. H.L. performed the experiments. H.L. and K.A.T. analyzed the data. M.J.Y. and Y.F. contributed to the development of the project. All authors contributed to the interpretation of the results. H.L. and K.A.T. wrote the manuscript, and all authors revised or commented.

\section{Preprints}
A preprint of this article is published at \\DOI:10.48550/arXiv.2205.10436.

\section{Data availability}
The data that support the findings of this study, as well as relevant
microscope images and scripts, have been deposited in Zenodo at
\verb|https://doi.org/10.5281/zenodo.11522483|.

%\subsection*{Code availability}
%The codes used to analyze the data are available upon request.

%\section{Material availability}
%The materials used in this work are available upon request.

\section{Competing interest statement}
There is no competing interest to declare.

\bibliography{ref}

\begin{thebibliography}{48}%
\makeatletter
\providecommand \@ifxundefined [1]{%
 \@ifx{#1\undefined}
}%
\providecommand \@ifnum [1]{%
 \ifnum #1\expandafter \@firstoftwo
 \else \expandafter \@secondoftwo
 \fi
}%
\providecommand \@ifx [1]{%
 \ifx #1\expandafter \@firstoftwo
 \else \expandafter \@secondoftwo
 \fi
}%
\providecommand \natexlab [1]{#1}%
\providecommand \enquote  [1]{``#1''}%
\providecommand \bibnamefont  [1]{#1}%
\providecommand \bibfnamefont [1]{#1}%
\providecommand \citenamefont [1]{#1}%
\providecommand \href@noop [0]{\@secondoftwo}%
\providecommand \href [0]{\begingroup \@sanitize@url \@href}%
\providecommand \@href[1]{\@@startlink{#1}\@@href}%
\providecommand \@@href[1]{\endgroup#1\@@endlink}%
\providecommand \@sanitize@url [0]{\catcode `\\12\catcode `\$12\catcode `\&12\catcode `\#12\catcode `\^12\catcode `\_12\catcode `\%12\relax}%
\providecommand \@@startlink[1]{}%
\providecommand \@@endlink[0]{}%
\providecommand \url  [0]{\begingroup\@sanitize@url \@url }%
\providecommand \@url [1]{\endgroup\@href {#1}{\urlprefix }}%
\providecommand \urlprefix  [0]{URL }%
\providecommand \Eprint [0]{\href }%
\providecommand \doibase [0]{http://dx.doi.org/}%
\providecommand \selectlanguage [0]{\@gobble}%
\providecommand \bibinfo  [0]{\@secondoftwo}%
\providecommand \bibfield  [0]{\@secondoftwo}%
\providecommand \translation [1]{[#1]}%
\providecommand \BibitemOpen [0]{}%
\providecommand \bibitemStop [0]{}%
\providecommand \bibitemNoStop [0]{.\EOS\space}%
\providecommand \EOS [0]{\spacefactor3000\relax}%
\providecommand \BibitemShut  [1]{\csname bibitem#1\endcsname}%
\let\auto@bib@innerbib\@empty
%</preamble>
\bibitem [{\citenamefont {Flemming}\ and\ \citenamefont {Wuertz}(2019)}]{Flemming.Wuertz-NRM2019}%
  \BibitemOpen
  \bibfield  {author} {\bibinfo {author} {\bibfnamefont {H.-C.}\ \bibnamefont {Flemming}}\ and\ \bibinfo {author} {\bibfnamefont {S.}~\bibnamefont {Wuertz}},\ }\href {\doibase 10.1038/s41579-019-0158-9} {\bibfield  {journal} {\bibinfo  {journal} {Nat. Rev. Microbiol.}\ }\textbf {\bibinfo {volume} {17}},\ \bibinfo {pages} {247} (\bibinfo {year} {2019})}\BibitemShut {NoStop}%
\bibitem [{\citenamefont {Sauer}\ \emph {et~al.}(2022)\citenamefont {Sauer}, \citenamefont {Stoodley}, \citenamefont {Goeres}, \citenamefont {Hall-Stoodley}, \citenamefont {Burm{\o}lle}, \citenamefont {Stewart},\ and\ \citenamefont {Bjarnsholt}}]{Sauer.etal-NRM2022}%
  \BibitemOpen
  \bibfield  {author} {\bibinfo {author} {\bibfnamefont {K.}~\bibnamefont {Sauer}}, \bibinfo {author} {\bibfnamefont {P.}~\bibnamefont {Stoodley}}, \bibinfo {author} {\bibfnamefont {D.~M.}\ \bibnamefont {Goeres}}, \bibinfo {author} {\bibfnamefont {L.}~\bibnamefont {Hall-Stoodley}}, \bibinfo {author} {\bibfnamefont {M.}~\bibnamefont {Burm{\o}lle}}, \bibinfo {author} {\bibfnamefont {P.~S.}\ \bibnamefont {Stewart}}, \ and\ \bibinfo {author} {\bibfnamefont {T.}~\bibnamefont {Bjarnsholt}},\ }\href {\doibase 10.1038/s41579-022-00767-0} {\bibfield  {journal} {\bibinfo  {journal} {Nat. Rev. Microbiol.}\ }\textbf {\bibinfo {volume} {20}},\ \bibinfo {pages} {608} (\bibinfo {year} {2022})}\BibitemShut {NoStop}%
\bibitem [{\citenamefont {Hall-Stoodley}\ \emph {et~al.}(2004)\citenamefont {Hall-Stoodley}, \citenamefont {Costerton},\ and\ \citenamefont {Stoodley}}]{HallStoodley.etal-NRM2004}%
  \BibitemOpen
  \bibfield  {author} {\bibinfo {author} {\bibfnamefont {L.}~\bibnamefont {Hall-Stoodley}}, \bibinfo {author} {\bibfnamefont {J.~W.}\ \bibnamefont {Costerton}}, \ and\ \bibinfo {author} {\bibfnamefont {P.}~\bibnamefont {Stoodley}},\ }\href {http://dx.doi.org/10.1038/nrmicro821} {\bibfield  {journal} {\bibinfo  {journal} {Nat. Rev. Microbiol.}\ }\textbf {\bibinfo {volume} {2}},\ \bibinfo {pages} {95} (\bibinfo {year} {2004})}\BibitemShut {NoStop}%
\bibitem [{\citenamefont {Vishwakarma}(2019)}]{Vishwakarma-JBM2019}%
  \BibitemOpen
  \bibfield  {author} {\bibinfo {author} {\bibfnamefont {V.}~\bibnamefont {Vishwakarma}},\ }\href {\doibase 10.1002/jobm.201900569} {\bibfield  {journal} {\bibinfo  {journal} {J. Basic Microbiol.}\ }\textbf {\bibinfo {volume} {60}},\ \bibinfo {pages} {198} (\bibinfo {year} {2019})}\BibitemShut {NoStop}%
\bibitem [{\citenamefont {Be'er}\ \emph {et~al.}(2020{\natexlab{a}})\citenamefont {Be'er}, \citenamefont {Ilkanaiv}, \citenamefont {Gross}, \citenamefont {Kearns}, \citenamefont {Heidenreich}, \citenamefont {Bär},\ and\ \citenamefont {Ariel}}]{Beer.etal-CP2020}%
  \BibitemOpen
  \bibfield  {author} {\bibinfo {author} {\bibfnamefont {A.}~\bibnamefont {Be'er}}, \bibinfo {author} {\bibfnamefont {B.}~\bibnamefont {Ilkanaiv}}, \bibinfo {author} {\bibfnamefont {R.}~\bibnamefont {Gross}}, \bibinfo {author} {\bibfnamefont {D.~B.}\ \bibnamefont {Kearns}}, \bibinfo {author} {\bibfnamefont {S.}~\bibnamefont {Heidenreich}}, \bibinfo {author} {\bibfnamefont {M.}~\bibnamefont {Bär}}, \ and\ \bibinfo {author} {\bibfnamefont {G.}~\bibnamefont {Ariel}},\ }\href {\doibase 10.1038/s42005-020-0327-1} {\bibfield  {journal} {\bibinfo  {journal} {Commun. Phys.}\ }\textbf {\bibinfo {volume} {3}},\ \bibinfo {pages} {66} (\bibinfo {year} {2020}{\natexlab{a}})}\BibitemShut {NoStop}%
\bibitem [{\citenamefont {Aranson}(2022)}]{Aranson-RPP2022}%
  \BibitemOpen
  \bibfield  {author} {\bibinfo {author} {\bibfnamefont {I.}~\bibnamefont {Aranson}},\ }\href {\doibase 10.1088/1361-6633/ac723d} {\bibfield  {journal} {\bibinfo  {journal} {Rep. Prog. Phys.}\ }\textbf {\bibinfo {volume} {85}},\ \bibinfo {pages} {076601} (\bibinfo {year} {2022})}\BibitemShut {NoStop}%
\bibitem [{\citenamefont {G\"otze}(2009)}]{Gotze-book}%
  \BibitemOpen
  \bibfield  {author} {\bibinfo {author} {\bibfnamefont {W.}~\bibnamefont {G\"otze}},\ }\href {\doibase 10.1093/acprof:oso/9780199235346.001.0001} {\emph {\bibinfo {title} {Complex Dynamics of Glass-Forming Liquids: A Mode-Coupling Theory}}},\ International series of monographs on physics\ (\bibinfo  {publisher} {Oxford Univ. Press},\ \bibinfo {address} {New York},\ \bibinfo {year} {2009})\BibitemShut {NoStop}%
\bibitem [{\citenamefont {Berthier}\ and\ \citenamefont {Biroli}(2011)}]{berthier2011theoretical}%
  \BibitemOpen
  \bibfield  {author} {\bibinfo {author} {\bibfnamefont {L.}~\bibnamefont {Berthier}}\ and\ \bibinfo {author} {\bibfnamefont {G.}~\bibnamefont {Biroli}},\ }\href@noop {} {\bibfield  {journal} {\bibinfo  {journal} {Rev. Mod. Phys.}\ }\textbf {\bibinfo {volume} {83}},\ \bibinfo {pages} {587} (\bibinfo {year} {2011})}\BibitemShut {NoStop}%
\bibitem [{\citenamefont {Hunter}\ and\ \citenamefont {Weeks}(2012)}]{hunter2012physics}%
  \BibitemOpen
  \bibfield  {author} {\bibinfo {author} {\bibfnamefont {G.~L.}\ \bibnamefont {Hunter}}\ and\ \bibinfo {author} {\bibfnamefont {E.~R.}\ \bibnamefont {Weeks}},\ }\href@noop {} {\bibfield  {journal} {\bibinfo  {journal} {Rep. Prog. Phys.}\ }\textbf {\bibinfo {volume} {75}},\ \bibinfo {pages} {066501} (\bibinfo {year} {2012})}\BibitemShut {NoStop}%
\bibitem [{\citenamefont {Charbonneau}\ \emph {et~al.}(2017)\citenamefont {Charbonneau}, \citenamefont {Kurchan}, \citenamefont {Parisi}, \citenamefont {Urbani},\ and\ \citenamefont {Zamponi}}]{Charbonneau.etal-ARCMP2017}%
  \BibitemOpen
  \bibfield  {author} {\bibinfo {author} {\bibfnamefont {P.}~\bibnamefont {Charbonneau}}, \bibinfo {author} {\bibfnamefont {J.}~\bibnamefont {Kurchan}}, \bibinfo {author} {\bibfnamefont {G.}~\bibnamefont {Parisi}}, \bibinfo {author} {\bibfnamefont {P.}~\bibnamefont {Urbani}}, \ and\ \bibinfo {author} {\bibfnamefont {F.}~\bibnamefont {Zamponi}},\ }\href {\doibase 10.1146/annurev-conmatphys-031016-025334} {\bibfield  {journal} {\bibinfo  {journal} {Annu. Rev. Condens. Matter Phys.}\ }\textbf {\bibinfo {volume} {8}},\ \bibinfo {pages} {265} (\bibinfo {year} {2017})}\BibitemShut {NoStop}%
\bibitem [{\citenamefont {Reichman}\ and\ \citenamefont {Charbonneau}(2005)}]{Reichman.Charbonneau-JSM2005}%
  \BibitemOpen
  \bibfield  {author} {\bibinfo {author} {\bibfnamefont {D.~R.}\ \bibnamefont {Reichman}}\ and\ \bibinfo {author} {\bibfnamefont {P.}~\bibnamefont {Charbonneau}},\ }\href {\doibase 10.1088/1742-5468/2005/05/p05013} {\bibfield  {journal} {\bibinfo  {journal} {J. Stat. Mech.}\ }\textbf {\bibinfo {volume} {2005}},\ \bibinfo {pages} {P05013} (\bibinfo {year} {2005})}\BibitemShut {NoStop}%
\bibitem [{\citenamefont {Parisi}\ \emph {et~al.}(2019)\citenamefont {Parisi}, \citenamefont {Urbani},\ and\ \citenamefont {Zamponi}}]{Parisi.etal-book2019}%
  \BibitemOpen
  \bibfield  {author} {\bibinfo {author} {\bibfnamefont {G.}~\bibnamefont {Parisi}}, \bibinfo {author} {\bibfnamefont {P.}~\bibnamefont {Urbani}}, \ and\ \bibinfo {author} {\bibfnamefont {F.}~\bibnamefont {Zamponi}},\ }\href {\doibase 10.1017/9781108120494} {\emph {\bibinfo {title} {Theory of Simple Glasses}}}\ (\bibinfo  {publisher} {Cambridge Univ. Press},\ \bibinfo {year} {2019})\BibitemShut {NoStop}%
\bibitem [{\citenamefont {Janssen}(2019)}]{janssen2019active}%
  \BibitemOpen
  \bibfield  {author} {\bibinfo {author} {\bibfnamefont {L.~M.~C.}\ \bibnamefont {Janssen}},\ }\href@noop {} {\bibfield  {journal} {\bibinfo  {journal} {J. Phys. Condens. Mat.}\ }\textbf {\bibinfo {volume} {31}},\ \bibinfo {pages} {503002} (\bibinfo {year} {2019})}\BibitemShut {NoStop}%
\bibitem [{\citenamefont {Berthier}\ \emph {et~al.}(2019)\citenamefont {Berthier}, \citenamefont {Flenner},\ and\ \citenamefont {Szamel}}]{Berthier.etal-JCP2019}%
  \BibitemOpen
  \bibfield  {author} {\bibinfo {author} {\bibfnamefont {L.}~\bibnamefont {Berthier}}, \bibinfo {author} {\bibfnamefont {E.}~\bibnamefont {Flenner}}, \ and\ \bibinfo {author} {\bibfnamefont {G.}~\bibnamefont {Szamel}},\ }\href {\doibase 10.1063/1.5093240} {\bibfield  {journal} {\bibinfo  {journal} {J. Chem. Phys.}\ }\textbf {\bibinfo {volume} {150}},\ \bibinfo {pages} {200901} (\bibinfo {year} {2019})}\BibitemShut {NoStop}%
\bibitem [{\citenamefont {Sadhukhan}\ \emph {et~al.}(2024)\citenamefont {Sadhukhan}, \citenamefont {Dey}, \citenamefont {Karmakar},\ and\ \citenamefont {Nandi}}]{Sadhukhan.etal-a2024}%
  \BibitemOpen
  \bibfield  {author} {\bibinfo {author} {\bibfnamefont {S.}~\bibnamefont {Sadhukhan}}, \bibinfo {author} {\bibfnamefont {S.}~\bibnamefont {Dey}}, \bibinfo {author} {\bibfnamefont {S.}~\bibnamefont {Karmakar}}, \ and\ \bibinfo {author} {\bibfnamefont {S.~K.}\ \bibnamefont {Nandi}},\ }\href {\doibase 10.1140/epjs/s11734-024-01188-1} {\bibfield  {journal} {\bibinfo  {journal} {Eur. Phys. J. Special Topics}\ } (\bibinfo {year} {2024}),\ 10.1140/epjs/s11734-024-01188-1},\ \Eprint {http://arxiv.org/abs/2403.06799} {arXiv:2403.06799} \BibitemShut {NoStop}%
\bibitem [{\citenamefont {Zhou}\ \emph {et~al.}(2009)\citenamefont {Zhou}, \citenamefont {Trepat}, \citenamefont {Park}, \citenamefont {Lenormand}, \citenamefont {Oliver}, \citenamefont {Mijailovich}, \citenamefont {Hardin}, \citenamefont {Weitz}, \citenamefont {Butler},\ and\ \citenamefont {Fredberg}}]{Zhou.etal-PNAS2009}%
  \BibitemOpen
  \bibfield  {author} {\bibinfo {author} {\bibfnamefont {E.~H.}\ \bibnamefont {Zhou}}, \bibinfo {author} {\bibfnamefont {X.}~\bibnamefont {Trepat}}, \bibinfo {author} {\bibfnamefont {C.~Y.}\ \bibnamefont {Park}}, \bibinfo {author} {\bibfnamefont {G.}~\bibnamefont {Lenormand}}, \bibinfo {author} {\bibfnamefont {M.~N.}\ \bibnamefont {Oliver}}, \bibinfo {author} {\bibfnamefont {S.~M.}\ \bibnamefont {Mijailovich}}, \bibinfo {author} {\bibfnamefont {C.}~\bibnamefont {Hardin}}, \bibinfo {author} {\bibfnamefont {D.~A.}\ \bibnamefont {Weitz}}, \bibinfo {author} {\bibfnamefont {J.~P.}\ \bibnamefont {Butler}}, \ and\ \bibinfo {author} {\bibfnamefont {J.~J.}\ \bibnamefont {Fredberg}},\ }\href {\doibase 10.1073/pnas.0901462106} {\bibfield  {journal} {\bibinfo  {journal} {Proc. Natl. Acad. Sci. USA}\ }\textbf {\bibinfo {volume} {106}},\ \bibinfo {pages} {10632} (\bibinfo {year} {2009})}\BibitemShut {NoStop}%
\bibitem [{\citenamefont {Parry}\ \emph {et~al.}(2014)\citenamefont {Parry}, \citenamefont {Surovtsev}, \citenamefont {Cabeen}, \citenamefont {O'Hern}, \citenamefont {Dufresne},\ and\ \citenamefont {Jacobs-Wagner}}]{Parry.etal-C2014}%
  \BibitemOpen
  \bibfield  {author} {\bibinfo {author} {\bibfnamefont {B.~R.}\ \bibnamefont {Parry}}, \bibinfo {author} {\bibfnamefont {I.~V.}\ \bibnamefont {Surovtsev}}, \bibinfo {author} {\bibfnamefont {M.~T.}\ \bibnamefont {Cabeen}}, \bibinfo {author} {\bibfnamefont {C.~S.}\ \bibnamefont {O'Hern}}, \bibinfo {author} {\bibfnamefont {E.~R.}\ \bibnamefont {Dufresne}}, \ and\ \bibinfo {author} {\bibfnamefont {C.}~\bibnamefont {Jacobs-Wagner}},\ }\href {\doibase 10.1016/j.cell.2013.11.028} {\bibfield  {journal} {\bibinfo  {journal} {Cell}\ }\textbf {\bibinfo {volume} {156}},\ \bibinfo {pages} {183} (\bibinfo {year} {2014})}\BibitemShut {NoStop}%
\bibitem [{\citenamefont {Nishizawa}\ \emph {et~al.}(2017)\citenamefont {Nishizawa}, \citenamefont {Fujiwara}, \citenamefont {Ikenaga}, \citenamefont {Nakajo}, \citenamefont {Yanagisawa},\ and\ \citenamefont {Mizuno}}]{Nishizawa.etal-SR2017}%
  \BibitemOpen
  \bibfield  {author} {\bibinfo {author} {\bibfnamefont {K.}~\bibnamefont {Nishizawa}}, \bibinfo {author} {\bibfnamefont {K.}~\bibnamefont {Fujiwara}}, \bibinfo {author} {\bibfnamefont {M.}~\bibnamefont {Ikenaga}}, \bibinfo {author} {\bibfnamefont {N.}~\bibnamefont {Nakajo}}, \bibinfo {author} {\bibfnamefont {M.}~\bibnamefont {Yanagisawa}}, \ and\ \bibinfo {author} {\bibfnamefont {D.}~\bibnamefont {Mizuno}},\ }\href {https://doi.org/10.1038/s41598-017-14883-y} {\bibfield  {journal} {\bibinfo  {journal} {Sci. Rep.}\ }\textbf {\bibinfo {volume} {7}},\ \bibinfo {pages} {15143} (\bibinfo {year} {2017})}\BibitemShut {NoStop}%
\bibitem [{\citenamefont {Angelini}\ \emph {et~al.}(2011)\citenamefont {Angelini}, \citenamefont {Hannezo}, \citenamefont {Trepat}, \citenamefont {Marquez}, \citenamefont {Fredberg},\ and\ \citenamefont {Weitz}}]{angelini2011glass}%
  \BibitemOpen
  \bibfield  {author} {\bibinfo {author} {\bibfnamefont {T.~E.}\ \bibnamefont {Angelini}}, \bibinfo {author} {\bibfnamefont {E.}~\bibnamefont {Hannezo}}, \bibinfo {author} {\bibfnamefont {X.}~\bibnamefont {Trepat}}, \bibinfo {author} {\bibfnamefont {M.}~\bibnamefont {Marquez}}, \bibinfo {author} {\bibfnamefont {J.~J.}\ \bibnamefont {Fredberg}}, \ and\ \bibinfo {author} {\bibfnamefont {D.~A.}\ \bibnamefont {Weitz}},\ }\href@noop {} {\bibfield  {journal} {\bibinfo  {journal} {Proc. Natl. Acad. Sci. USA}\ }\textbf {\bibinfo {volume} {108}},\ \bibinfo {pages} {4714} (\bibinfo {year} {2011})}\BibitemShut {NoStop}%
\bibitem [{\citenamefont {Park}\ \emph {et~al.}(2015)\citenamefont {Park}, \citenamefont {Kim}, \citenamefont {Bi}, \citenamefont {Mitchel}, \citenamefont {Qazvini}, \citenamefont {Tantisira}, \citenamefont {Park}, \citenamefont {McGill}, \citenamefont {Kim}, \citenamefont {Gweon} \emph {et~al.}}]{park2015unjamming}%
  \BibitemOpen
  \bibfield  {author} {\bibinfo {author} {\bibfnamefont {J.-A.}\ \bibnamefont {Park}}, \bibinfo {author} {\bibfnamefont {J.~H.}\ \bibnamefont {Kim}}, \bibinfo {author} {\bibfnamefont {D.}~\bibnamefont {Bi}}, \bibinfo {author} {\bibfnamefont {J.~A.}\ \bibnamefont {Mitchel}}, \bibinfo {author} {\bibfnamefont {N.~T.}\ \bibnamefont {Qazvini}}, \bibinfo {author} {\bibfnamefont {K.}~\bibnamefont {Tantisira}}, \bibinfo {author} {\bibfnamefont {C.~Y.}\ \bibnamefont {Park}}, \bibinfo {author} {\bibfnamefont {M.}~\bibnamefont {McGill}}, \bibinfo {author} {\bibfnamefont {S.-H.}\ \bibnamefont {Kim}}, \bibinfo {author} {\bibfnamefont {B.}~\bibnamefont {Gweon}},  \emph {et~al.},\ }\href@noop {} {\bibfield  {journal} {\bibinfo  {journal} {Nat. Mater.}\ }\textbf {\bibinfo {volume} {14}},\ \bibinfo {pages} {1040} (\bibinfo {year} {2015})}\BibitemShut {NoStop}%
\bibitem [{\citenamefont {Garcia}\ \emph {et~al.}(2015)\citenamefont {Garcia}, \citenamefont {Hannezo}, \citenamefont {Elgeti}, \citenamefont {Joanny}, \citenamefont {Silberzan},\ and\ \citenamefont {Gov}}]{Garcia.etal-PNAS2015}%
  \BibitemOpen
  \bibfield  {author} {\bibinfo {author} {\bibfnamefont {S.}~\bibnamefont {Garcia}}, \bibinfo {author} {\bibfnamefont {E.}~\bibnamefont {Hannezo}}, \bibinfo {author} {\bibfnamefont {J.}~\bibnamefont {Elgeti}}, \bibinfo {author} {\bibfnamefont {J.-F.}\ \bibnamefont {Joanny}}, \bibinfo {author} {\bibfnamefont {P.}~\bibnamefont {Silberzan}}, \ and\ \bibinfo {author} {\bibfnamefont {N.~S.}\ \bibnamefont {Gov}},\ }\href {\doibase 10.1073/pnas.1510973112} {\bibfield  {journal} {\bibinfo  {journal} {Proc. Natl. Acad. Sci. USA}\ }\textbf {\bibinfo {volume} {112}},\ \bibinfo {pages} {15314} (\bibinfo {year} {2015})}\BibitemShut {NoStop}%
\bibitem [{\citenamefont {Rhodeland}\ \emph {et~al.}(2020)\citenamefont {Rhodeland}, \citenamefont {Hoeger},\ and\ \citenamefont {Ursell}}]{Rhodeland.etal-JRSI2020}%
  \BibitemOpen
  \bibfield  {author} {\bibinfo {author} {\bibfnamefont {B.}~\bibnamefont {Rhodeland}}, \bibinfo {author} {\bibfnamefont {K.}~\bibnamefont {Hoeger}}, \ and\ \bibinfo {author} {\bibfnamefont {T.}~\bibnamefont {Ursell}},\ }\href {\doibase 10.1098/rsif.2020.0147} {\bibfield  {journal} {\bibinfo  {journal} {J. R. Soc. Interface}\ }\textbf {\bibinfo {volume} {17}},\ \bibinfo {pages} {20200147} (\bibinfo {year} {2020})}\BibitemShut {NoStop}%
\bibitem [{\citenamefont {Takatori}\ and\ \citenamefont {Mandadapu}(2020)}]{takatori2020motility}%
  \BibitemOpen
  \bibfield  {author} {\bibinfo {author} {\bibfnamefont {S.~C.}\ \bibnamefont {Takatori}}\ and\ \bibinfo {author} {\bibfnamefont {K.~K.}\ \bibnamefont {Mandadapu}},\ }\href@noop {} {\bibfield  {journal} {\bibinfo  {journal} {arXiv:2003.05618}\ } (\bibinfo {year} {2020})}\BibitemShut {NoStop}%
\bibitem [{\citenamefont {Mather}\ \emph {et~al.}(2010)\citenamefont {Mather}, \citenamefont {Mondrag\'on-Palomino}, \citenamefont {Danino}, \citenamefont {Hasty},\ and\ \citenamefont {Tsimring}}]{Mather.etal-PRL2010}%
  \BibitemOpen
  \bibfield  {author} {\bibinfo {author} {\bibfnamefont {W.}~\bibnamefont {Mather}}, \bibinfo {author} {\bibfnamefont {O.}~\bibnamefont {Mondrag\'on-Palomino}}, \bibinfo {author} {\bibfnamefont {T.}~\bibnamefont {Danino}}, \bibinfo {author} {\bibfnamefont {J.}~\bibnamefont {Hasty}}, \ and\ \bibinfo {author} {\bibfnamefont {L.~S.}\ \bibnamefont {Tsimring}},\ }\href {\doibase 10.1103/PhysRevLett.104.208101} {\bibfield  {journal} {\bibinfo  {journal} {Phys. Rev. Lett.}\ }\textbf {\bibinfo {volume} {104}},\ \bibinfo {pages} {208101} (\bibinfo {year} {2010})}\BibitemShut {NoStop}%
\bibitem [{\citenamefont {Shimaya}\ \emph {et~al.}(2021)\citenamefont {Shimaya}, \citenamefont {Okura}, \citenamefont {Wakamoto},\ and\ \citenamefont {Takeuchi}}]{Shimaya2021}%
  \BibitemOpen
  \bibfield  {author} {\bibinfo {author} {\bibfnamefont {T.}~\bibnamefont {Shimaya}}, \bibinfo {author} {\bibfnamefont {R.}~\bibnamefont {Okura}}, \bibinfo {author} {\bibfnamefont {Y.}~\bibnamefont {Wakamoto}}, \ and\ \bibinfo {author} {\bibfnamefont {K.~A.}\ \bibnamefont {Takeuchi}},\ }\href@noop {} {\bibfield  {journal} {\bibinfo  {journal} {Commun. Phys.}\ }\textbf {\bibinfo {volume} {4}},\ \bibinfo {pages} {238} (\bibinfo {year} {2021})}\BibitemShut {NoStop}%
\bibitem [{\citenamefont {Pastore}\ \emph {et~al.}(2017)\citenamefont {Pastore}, \citenamefont {Pesce},\ and\ \citenamefont {Caggioni}}]{pastore2017differential}%
  \BibitemOpen
  \bibfield  {author} {\bibinfo {author} {\bibfnamefont {R.}~\bibnamefont {Pastore}}, \bibinfo {author} {\bibfnamefont {G.}~\bibnamefont {Pesce}}, \ and\ \bibinfo {author} {\bibfnamefont {M.}~\bibnamefont {Caggioni}},\ }\href {\doibase 10.1038/srep43496} {\bibfield  {journal} {\bibinfo  {journal} {Sci. Rep.}\ }\textbf {\bibinfo {volume} {7}},\ \bibinfo {pages} {43496} (\bibinfo {year} {2017})}\BibitemShut {NoStop}%
\bibitem [{\citenamefont {You}\ \emph {et~al.}(2018)\citenamefont {You}, \citenamefont {Pearce}, \citenamefont {Sengupta},\ and\ \citenamefont {Giomi}}]{You.etal-PRX2018}%
  \BibitemOpen
  \bibfield  {author} {\bibinfo {author} {\bibfnamefont {Z.}~\bibnamefont {You}}, \bibinfo {author} {\bibfnamefont {D.~J.~G.}\ \bibnamefont {Pearce}}, \bibinfo {author} {\bibfnamefont {A.}~\bibnamefont {Sengupta}}, \ and\ \bibinfo {author} {\bibfnamefont {L.}~\bibnamefont {Giomi}},\ }\href {\doibase 10.1103/PhysRevX.8.031065} {\bibfield  {journal} {\bibinfo  {journal} {Phys. Rev. X}\ }\textbf {\bibinfo {volume} {8}},\ \bibinfo {pages} {031065} (\bibinfo {year} {2018})}\BibitemShut {NoStop}%
\bibitem [{\citenamefont {Marchetti}\ \emph {et~al.}(2013)\citenamefont {Marchetti}, \citenamefont {Joanny}, \citenamefont {Ramaswamy}, \citenamefont {Liverpool}, \citenamefont {Prost}, \citenamefont {Rao},\ and\ \citenamefont {Simha}}]{marchetti2013hydrodynamics}%
  \BibitemOpen
  \bibfield  {author} {\bibinfo {author} {\bibfnamefont {M.~C.}\ \bibnamefont {Marchetti}}, \bibinfo {author} {\bibfnamefont {J.-F.}\ \bibnamefont {Joanny}}, \bibinfo {author} {\bibfnamefont {S.}~\bibnamefont {Ramaswamy}}, \bibinfo {author} {\bibfnamefont {T.~B.}\ \bibnamefont {Liverpool}}, \bibinfo {author} {\bibfnamefont {J.}~\bibnamefont {Prost}}, \bibinfo {author} {\bibfnamefont {M.}~\bibnamefont {Rao}}, \ and\ \bibinfo {author} {\bibfnamefont {R.~A.}\ \bibnamefont {Simha}},\ }\href@noop {} {\bibfield  {journal} {\bibinfo  {journal} {Rev. Mod. Phys.}\ }\textbf {\bibinfo {volume} {85}},\ \bibinfo {pages} {1143} (\bibinfo {year} {2013})}\BibitemShut {NoStop}%
\bibitem [{\citenamefont {Bechinger}\ \emph {et~al.}(2016)\citenamefont {Bechinger}, \citenamefont {Di~Leonardo}, \citenamefont {L{\"o}wen}, \citenamefont {Reichhardt}, \citenamefont {Volpe},\ and\ \citenamefont {Volpe}}]{bechinger2016active}%
  \BibitemOpen
  \bibfield  {author} {\bibinfo {author} {\bibfnamefont {C.}~\bibnamefont {Bechinger}}, \bibinfo {author} {\bibfnamefont {R.}~\bibnamefont {Di~Leonardo}}, \bibinfo {author} {\bibfnamefont {H.}~\bibnamefont {L{\"o}wen}}, \bibinfo {author} {\bibfnamefont {C.}~\bibnamefont {Reichhardt}}, \bibinfo {author} {\bibfnamefont {G.}~\bibnamefont {Volpe}}, \ and\ \bibinfo {author} {\bibfnamefont {G.}~\bibnamefont {Volpe}},\ }\href@noop {} {\bibfield  {journal} {\bibinfo  {journal} {Rev. Mod. Phys.}\ }\textbf {\bibinfo {volume} {88}},\ \bibinfo {pages} {045006} (\bibinfo {year} {2016})}\BibitemShut {NoStop}%
\bibitem [{\citenamefont {Doostmohammadi}\ and\ \citenamefont {Yeomans}(2019)}]{doostmohammadi2019coherent}%
  \BibitemOpen
  \bibfield  {author} {\bibinfo {author} {\bibfnamefont {A.}~\bibnamefont {Doostmohammadi}}\ and\ \bibinfo {author} {\bibfnamefont {J.~M.}\ \bibnamefont {Yeomans}},\ }\href@noop {} {\bibfield  {journal} {\bibinfo  {journal} {Eur. Phys. J. Special Topics}\ }\textbf {\bibinfo {volume} {227}},\ \bibinfo {pages} {2401} (\bibinfo {year} {2019})}\BibitemShut {NoStop}%
\bibitem [{\citenamefont {Be'er}\ and\ \citenamefont {Ariel}(2019)}]{be2019statistical}%
  \BibitemOpen
  \bibfield  {author} {\bibinfo {author} {\bibfnamefont {A.}~\bibnamefont {Be'er}}\ and\ \bibinfo {author} {\bibfnamefont {G.}~\bibnamefont {Ariel}},\ }\href {\doibase 10.1186/s40462-019-0153-9} {\bibfield  {journal} {\bibinfo  {journal} {Mov. Ecol.}\ }\textbf {\bibinfo {volume} {7}},\ \bibinfo {pages} {9} (\bibinfo {year} {2019})}\BibitemShut {NoStop}%
\bibitem [{\citenamefont {Be'er}\ \emph {et~al.}(2020{\natexlab{b}})\citenamefont {Be'er}, \citenamefont {Ilkanaiv}, \citenamefont {Gross}, \citenamefont {Kearns}, \citenamefont {Heidenreich}, \citenamefont {Bär},\ and\ \citenamefont {Ariel}}]{be2020phase}%
  \BibitemOpen
  \bibfield  {author} {\bibinfo {author} {\bibfnamefont {A.}~\bibnamefont {Be'er}}, \bibinfo {author} {\bibfnamefont {B.}~\bibnamefont {Ilkanaiv}}, \bibinfo {author} {\bibfnamefont {R.}~\bibnamefont {Gross}}, \bibinfo {author} {\bibfnamefont {D.~B.}\ \bibnamefont {Kearns}}, \bibinfo {author} {\bibfnamefont {S.}~\bibnamefont {Heidenreich}}, \bibinfo {author} {\bibfnamefont {M.}~\bibnamefont {Bär}}, \ and\ \bibinfo {author} {\bibfnamefont {G.}~\bibnamefont {Ariel}},\ }\href {\doibase 10.1038/s42005-020-0327-1} {\bibfield  {journal} {\bibinfo  {journal} {Commun. Phys.}\ }\textbf {\bibinfo {volume} {3}},\ \bibinfo {pages} {66} (\bibinfo {year} {2020}{\natexlab{b}})}\BibitemShut {NoStop}%
\bibitem [{\citenamefont {Szamel}\ \emph {et~al.}(2015)\citenamefont {Szamel}, \citenamefont {Flenner},\ and\ \citenamefont {Berthier}}]{Szamel.etal-PRE2015}%
  \BibitemOpen
  \bibfield  {author} {\bibinfo {author} {\bibfnamefont {G.}~\bibnamefont {Szamel}}, \bibinfo {author} {\bibfnamefont {E.}~\bibnamefont {Flenner}}, \ and\ \bibinfo {author} {\bibfnamefont {L.}~\bibnamefont {Berthier}},\ }\href {\doibase 10.1103/PhysRevE.91.062304} {\bibfield  {journal} {\bibinfo  {journal} {Phys. Rev. E}\ }\textbf {\bibinfo {volume} {91}},\ \bibinfo {pages} {062304} (\bibinfo {year} {2015})}\BibitemShut {NoStop}%
\bibitem [{\citenamefont {Dey}\ \emph {et~al.}(2022)\citenamefont {Dey}, \citenamefont {Mutneja},\ and\ \citenamefont {Karmakar}}]{Dey.etal-SM2022}%
  \BibitemOpen
  \bibfield  {author} {\bibinfo {author} {\bibfnamefont {S.}~\bibnamefont {Dey}}, \bibinfo {author} {\bibfnamefont {A.}~\bibnamefont {Mutneja}}, \ and\ \bibinfo {author} {\bibfnamefont {S.}~\bibnamefont {Karmakar}},\ }\href {\doibase 10.1039/d2sm00727d} {\bibfield  {journal} {\bibinfo  {journal} {Soft Matter}\ }\textbf {\bibinfo {volume} {18}},\ \bibinfo {pages} {7309} (\bibinfo {year} {2022})}\BibitemShut {NoStop}%
\bibitem [{\citenamefont {Zheng}\ \emph {et~al.}(2011)\citenamefont {Zheng}, \citenamefont {Wang}, \citenamefont {Han} \emph {et~al.}}]{zheng2011glass}%
  \BibitemOpen
  \bibfield  {author} {\bibinfo {author} {\bibfnamefont {Z.}~\bibnamefont {Zheng}}, \bibinfo {author} {\bibfnamefont {F.}~\bibnamefont {Wang}}, \bibinfo {author} {\bibfnamefont {Y.}~\bibnamefont {Han}},  \emph {et~al.},\ }\href@noop {} {\bibfield  {journal} {\bibinfo  {journal} {Phys. Rev Lett}\ }\textbf {\bibinfo {volume} {107}},\ \bibinfo {pages} {065702} (\bibinfo {year} {2011})}\BibitemShut {NoStop}%
\bibitem [{\citenamefont {Mishra}\ \emph {et~al.}(2013)\citenamefont {Mishra}, \citenamefont {Rangarajan},\ and\ \citenamefont {Ganapathy}}]{Mishra.etal-PRL2013}%
  \BibitemOpen
  \bibfield  {author} {\bibinfo {author} {\bibfnamefont {C.~K.}\ \bibnamefont {Mishra}}, \bibinfo {author} {\bibfnamefont {A.}~\bibnamefont {Rangarajan}}, \ and\ \bibinfo {author} {\bibfnamefont {R.}~\bibnamefont {Ganapathy}},\ }\href {\doibase 10.1103/PhysRevLett.110.188301} {\bibfield  {journal} {\bibinfo  {journal} {Phys. Rev. Lett.}\ }\textbf {\bibinfo {volume} {110}},\ \bibinfo {pages} {188301} (\bibinfo {year} {2013})}\BibitemShut {NoStop}%
\bibitem [{\citenamefont {Zheng}\ \emph {et~al.}(2014)\citenamefont {Zheng}, \citenamefont {Ni}, \citenamefont {Wang}, \citenamefont {Dijkstra}, \citenamefont {Wang},\ and\ \citenamefont {Han}}]{zheng2014structural}%
  \BibitemOpen
  \bibfield  {author} {\bibinfo {author} {\bibfnamefont {Z.}~\bibnamefont {Zheng}}, \bibinfo {author} {\bibfnamefont {R.}~\bibnamefont {Ni}}, \bibinfo {author} {\bibfnamefont {F.}~\bibnamefont {Wang}}, \bibinfo {author} {\bibfnamefont {M.}~\bibnamefont {Dijkstra}}, \bibinfo {author} {\bibfnamefont {Y.}~\bibnamefont {Wang}}, \ and\ \bibinfo {author} {\bibfnamefont {Y.}~\bibnamefont {Han}},\ }\href {\doibase 10.1038/ncomms4829} {\bibfield  {journal} {\bibinfo  {journal} {Nat. Commun.}\ }\textbf {\bibinfo {volume} {5}},\ \bibinfo {pages} {3829} (\bibinfo {year} {2014})}\BibitemShut {NoStop}%
\bibitem [{\citenamefont {Roller}\ \emph {et~al.}(2021)\citenamefont {Roller}, \citenamefont {Laganapan}, \citenamefont {Meijer}, \citenamefont {Fuchs},\ and\ \citenamefont {Zumbusch}}]{Roller.etal-PNAS2021}%
  \BibitemOpen
  \bibfield  {author} {\bibinfo {author} {\bibfnamefont {J.}~\bibnamefont {Roller}}, \bibinfo {author} {\bibfnamefont {A.}~\bibnamefont {Laganapan}}, \bibinfo {author} {\bibfnamefont {J.-M.}\ \bibnamefont {Meijer}}, \bibinfo {author} {\bibfnamefont {M.}~\bibnamefont {Fuchs}}, \ and\ \bibinfo {author} {\bibfnamefont {A.}~\bibnamefont {Zumbusch}},\ }\href {\doibase 10.1073/pnas.2018072118} {\bibfield  {journal} {\bibinfo  {journal} {Proc. Natl. Acad. Sci. USA}\ }\textbf {\bibinfo {volume} {118}},\ \bibinfo {pages} {e2018072118} (\bibinfo {year} {2021})}\BibitemShut {NoStop}%
\bibitem [{\citenamefont {Letz}\ \emph {et~al.}(2000)\citenamefont {Letz}, \citenamefont {Schilling},\ and\ \citenamefont {Latz}}]{Letz.etal-PRE2000}%
  \BibitemOpen
  \bibfield  {author} {\bibinfo {author} {\bibfnamefont {M.}~\bibnamefont {Letz}}, \bibinfo {author} {\bibfnamefont {R.}~\bibnamefont {Schilling}}, \ and\ \bibinfo {author} {\bibfnamefont {A.}~\bibnamefont {Latz}},\ }\href {\doibase 10.1103/PhysRevE.62.5173} {\bibfield  {journal} {\bibinfo  {journal} {Phys. Rev. E}\ }\textbf {\bibinfo {volume} {62}},\ \bibinfo {pages} {5173} (\bibinfo {year} {2000})}\BibitemShut {NoStop}%
\bibitem [{\citenamefont {Mandal}\ \emph {et~al.}(2017)\citenamefont {Mandal}, \citenamefont {Bhuyan}, \citenamefont {Chaudhuri}, \citenamefont {Rao},\ and\ \citenamefont {Dasgupta}}]{Mandal.etal-PRE2017}%
  \BibitemOpen
  \bibfield  {author} {\bibinfo {author} {\bibfnamefont {R.}~\bibnamefont {Mandal}}, \bibinfo {author} {\bibfnamefont {P.~J.}\ \bibnamefont {Bhuyan}}, \bibinfo {author} {\bibfnamefont {P.}~\bibnamefont {Chaudhuri}}, \bibinfo {author} {\bibfnamefont {M.}~\bibnamefont {Rao}}, \ and\ \bibinfo {author} {\bibfnamefont {C.}~\bibnamefont {Dasgupta}},\ }\href {\doibase 10.1103/PhysRevE.96.042605} {\bibfield  {journal} {\bibinfo  {journal} {Phys. Rev. E}\ }\textbf {\bibinfo {volume} {96}},\ \bibinfo {pages} {042605} (\bibinfo {year} {2017})}\BibitemShut {NoStop}%
\bibitem [{\citenamefont {Panigrahi}\ \emph {et~al.}(2021)\citenamefont {Panigrahi}, \citenamefont {Murat}, \citenamefont {Gall}, \citenamefont {Martineau}, \citenamefont {Goldlust}, \citenamefont {Fiche}, \citenamefont {Rombouts}, \citenamefont {Nöllmann}, \citenamefont {Espinosa},\ and\ \citenamefont {Mignot}}]{Panigrahi.etal-e2021}%
  \BibitemOpen
  \bibfield  {author} {\bibinfo {author} {\bibfnamefont {S.}~\bibnamefont {Panigrahi}}, \bibinfo {author} {\bibfnamefont {D.}~\bibnamefont {Murat}}, \bibinfo {author} {\bibfnamefont {A.~L.}\ \bibnamefont {Gall}}, \bibinfo {author} {\bibfnamefont {E.}~\bibnamefont {Martineau}}, \bibinfo {author} {\bibfnamefont {K.}~\bibnamefont {Goldlust}}, \bibinfo {author} {\bibfnamefont {J.-B.}\ \bibnamefont {Fiche}}, \bibinfo {author} {\bibfnamefont {S.}~\bibnamefont {Rombouts}}, \bibinfo {author} {\bibfnamefont {M.}~\bibnamefont {Nöllmann}}, \bibinfo {author} {\bibfnamefont {L.}~\bibnamefont {Espinosa}}, \ and\ \bibinfo {author} {\bibfnamefont {T.}~\bibnamefont {Mignot}},\ }\href {\doibase 10.7554/elife.65151} {\bibfield  {journal} {\bibinfo  {journal} {{eLife}}\ }\textbf {\bibinfo {volume} {10}},\ \bibinfo {pages} {e65151} (\bibinfo {year} {2021})}\BibitemShut {NoStop}%
\bibitem [{\citenamefont {Pant}\ \emph {et~al.}(2020)\citenamefont {Pant}, \citenamefont {Sharma}, \citenamefont {Verma}, \citenamefont {Singla},\ and\ \citenamefont {Sikander}}]{FuzzyBook}%
  \BibitemOpen
  \bibfield  {author} {\bibinfo {author} {\bibfnamefont {M.}~\bibnamefont {Pant}}, \bibinfo {author} {\bibfnamefont {T.}~\bibnamefont {Sharma}}, \bibinfo {author} {\bibfnamefont {O.}~\bibnamefont {Verma}}, \bibinfo {author} {\bibfnamefont {R.}~\bibnamefont {Singla}}, \ and\ \bibinfo {author} {\bibfnamefont {A.}~\bibnamefont {Sikander}},\ }\href {https://books.google.co.jp/books?id=9aDSDwAAQBAJ} {\emph {\bibinfo {title} {Soft Computing: Theories and Applications: Proceedings of SoCTA 2018}}},\ Advances in Intelligent Systems and Computing\ (\bibinfo  {publisher} {Springer Singapore},\ \bibinfo {year} {2020})\BibitemShut {NoStop}%
\bibitem [{\citenamefont {Min}\ \emph {et~al.}(2009)\citenamefont {Min}, \citenamefont {Mears}, \citenamefont {Chubiz}, \citenamefont {Rao}, \citenamefont {Golding},\ and\ \citenamefont {Chemla}}]{Min.etal-NM2009}%
  \BibitemOpen
  \bibfield  {author} {\bibinfo {author} {\bibfnamefont {T.~L.}\ \bibnamefont {Min}}, \bibinfo {author} {\bibfnamefont {P.~J.}\ \bibnamefont {Mears}}, \bibinfo {author} {\bibfnamefont {L.~M.}\ \bibnamefont {Chubiz}}, \bibinfo {author} {\bibfnamefont {C.~V.}\ \bibnamefont {Rao}}, \bibinfo {author} {\bibfnamefont {I.}~\bibnamefont {Golding}}, \ and\ \bibinfo {author} {\bibfnamefont {Y.~R.}\ \bibnamefont {Chemla}},\ }\href {\doibase 10.1038/nmeth.1380} {\bibfield  {journal} {\bibinfo  {journal} {Nat. Meth.}\ }\textbf {\bibinfo {volume} {6}},\ \bibinfo {pages} {831} (\bibinfo {year} {2009})}\BibitemShut {NoStop}%
\bibitem [{\citenamefont {J{\"a}hne}(1993)}]{Jahne-book}%
  \BibitemOpen
  \bibfield  {author} {\bibinfo {author} {\bibfnamefont {B.}~\bibnamefont {J{\"a}hne}},\ }\href {https://books.google.fr/books?id=gO6V5gh4IXsC} {\emph {\bibinfo {title} {Spatio-Temporal Image Processing: Theory and Scientific Applications}}},\ Lecture Notes in Computer Science\ (\bibinfo  {publisher} {Springer Berlin Heidelberg},\ \bibinfo {year} {1993})\BibitemShut {NoStop}%
\bibitem [{\citenamefont {Crocker}\ and\ \citenamefont {Grier}(1996)}]{Crocker.Grier-JCIS1996}%
  \BibitemOpen
  \bibfield  {author} {\bibinfo {author} {\bibfnamefont {J.~C.}\ \bibnamefont {Crocker}}\ and\ \bibinfo {author} {\bibfnamefont {D.~G.}\ \bibnamefont {Grier}},\ }\href {\doibase 10.1006/jcis.1996.0217} {\bibfield  {journal} {\bibinfo  {journal} {J. Colloid Interface Sci.}\ }\textbf {\bibinfo {volume} {179}},\ \bibinfo {pages} {298} (\bibinfo {year} {1996})}\BibitemShut {NoStop}%
\bibitem [{\citenamefont {Schilling}\ and\ \citenamefont {Scheidsteger}(1997)}]{Schilling.Scheidsteger-PRE1997}%
  \BibitemOpen
  \bibfield  {author} {\bibinfo {author} {\bibfnamefont {R.}~\bibnamefont {Schilling}}\ and\ \bibinfo {author} {\bibfnamefont {T.}~\bibnamefont {Scheidsteger}},\ }\href {\doibase 10.1103/PhysRevE.56.2932} {\bibfield  {journal} {\bibinfo  {journal} {Phys. Rev. E}\ }\textbf {\bibinfo {volume} {56}},\ \bibinfo {pages} {2932} (\bibinfo {year} {1997})}\BibitemShut {NoStop}%
\bibitem [{\citenamefont {Moreno}\ \emph {et~al.}(2005)\citenamefont {Moreno}, \citenamefont {Chong}, \citenamefont {Kob},\ and\ \citenamefont {Sciortino}}]{Moreno.etal-JCP2005}%
  \BibitemOpen
  \bibfield  {author} {\bibinfo {author} {\bibfnamefont {A.~J.}\ \bibnamefont {Moreno}}, \bibinfo {author} {\bibfnamefont {S.-H.}\ \bibnamefont {Chong}}, \bibinfo {author} {\bibfnamefont {W.}~\bibnamefont {Kob}}, \ and\ \bibinfo {author} {\bibfnamefont {F.}~\bibnamefont {Sciortino}},\ }\href {\doibase 10.1063/1.2085030} {\bibfield  {journal} {\bibinfo  {journal} {J. Chem. Phys.}\ }\textbf {\bibinfo {volume} {123}},\ \bibinfo {pages} {204505} (\bibinfo {year} {2005})}\BibitemShut {NoStop}%
\bibitem [{\citenamefont {Pfleiderer}\ \emph {et~al.}(2008)\citenamefont {Pfleiderer}, \citenamefont {Milinkovic},\ and\ \citenamefont {Schilling}}]{Pfleiderer.etal-EL2008}%
  \BibitemOpen
  \bibfield  {author} {\bibinfo {author} {\bibfnamefont {P.}~\bibnamefont {Pfleiderer}}, \bibinfo {author} {\bibfnamefont {K.}~\bibnamefont {Milinkovic}}, \ and\ \bibinfo {author} {\bibfnamefont {T.}~\bibnamefont {Schilling}},\ }\href {\doibase 10.1209/0295-5075/84/16003} {\bibfield  {journal} {\bibinfo  {journal} {Europhys. Lett.}\ }\textbf {\bibinfo {volume} {84}},\ \bibinfo {pages} {16003} (\bibinfo {year} {2008})}\BibitemShut {NoStop}%
\end{thebibliography}%


\begin{thebibliography}{0}%
\makeatletter
\providecommand \@ifxundefined [1]{%
 \@ifx{#1\undefined}
}%
\providecommand \@ifnum [1]{%
 \ifnum #1\expandafter \@firstoftwo
 \else \expandafter \@secondoftwo
 \fi
}%
\providecommand \@ifx [1]{%
 \ifx #1\expandafter \@firstoftwo
 \else \expandafter \@secondoftwo
 \fi
}%
\providecommand \natexlab [1]{#1}%
\providecommand \enquote  [1]{``#1''}%
\providecommand \bibnamefont  [1]{#1}%
\providecommand \bibfnamefont [1]{#1}%
\providecommand \citenamefont [1]{#1}%
\providecommand \href@noop [0]{\@secondoftwo}%
\providecommand \href [0]{\begingroup \@sanitize@url \@href}%
\providecommand \@href[1]{\@@startlink{#1}\@@href}%
\providecommand \@@href[1]{\endgroup#1\@@endlink}%
\providecommand \@sanitize@url [0]{\catcode `\\12\catcode `\$12\catcode `\&12\catcode `\#12\catcode `\^12\catcode `\_12\catcode `\%12\relax}%
\providecommand \@@startlink[1]{}%
\providecommand \@@endlink[0]{}%
\providecommand \url  [0]{\begingroup\@sanitize@url \@url }%
\providecommand \@url [1]{\endgroup\@href {#1}{\urlprefix }}%
\providecommand \urlprefix  [0]{URL }%
\providecommand \Eprint [0]{\href }%
\providecommand \doibase [0]{http://dx.doi.org/}%
\providecommand \selectlanguage [0]{\@gobble}%
\providecommand \bibinfo  [0]{\@secondoftwo}%
\providecommand \bibfield  [0]{\@secondoftwo}%
\providecommand \translation [1]{[#1]}%
\providecommand \BibitemOpen [0]{}%
\providecommand \bibitemStop [0]{}%
\providecommand \bibitemNoStop [0]{.\EOS\space}%
\providecommand \EOS [0]{\spacefactor3000\relax}%
\providecommand \BibitemShut  [1]{\csname bibitem#1\endcsname}%
\let\auto@bib@innerbib\@empty
%</preamble>
\end{thebibliography}%

\section{Appendix A: Materials and Methods}

\subsection*{Culture of bacteria} 
We used a motile strain of \textit{Escherichia coli}, RP437. 
As one of the standard strains for motility studies, its run-and-tumble behavior is well documented in the literature \cite{Min.etal-NM2009}.
First we inoculated bacteria into a sterile test tube with $5\unit{mL}$ of Luria-Bertani (LB) medium (containing bacto-tryptone $10\unit{g/L}$, yeast extract $5\unit{g/L}$, and NaCl $10\unit{g/L}$). Then we incubated the liquid culture inside a shaking incubator at $37\unit{^\circ C}$ for $12\unit{h}$ at the shaking speed of $220\unit{rpm}$. We resuspended $100\unit{\mu L}$ of the overnight culture into $10\unit{mL}$ of fresh tryptone broth (TB, containing bacto-tryptone $10\unit{g/L}$ and NaCl $10\unit{g/L}$), and again incubated it for $4\unit{h}$ in the same culture condition. Finally, we measured the optical density of the liquid culture by a spectrophotometer. The optical density at wave length $600\unit{nm}, \mathrm{OD}_{600}$, was $0.4$, which is equivalent to $10^8\unit{cells/mL}$. We diluted the liquid culture to a concentration of $\mathrm{OD}_{600}=0.1$ ($\sim 10^7\unit{cells/mL}$) by resuspending it into a fresh TB medium containing $0.02\unit{wt\%}$ of surfactant Tween-20. 

\subsection*{Fabrication of microfluidic device} 
We performed experiments using a membrane-based microfluidic device, namely, the extensive microperfusion system (EMPS) \cite{Shimaya2021} (\figref[a]{fig1}). The EMPS comprises a micropatterned coverslip, a bilayer porous membrane (cellulose and polyethylene terephthalate), and a polydimethylsiloxane (PDMS) pad with an inlet and two outlets. The coverslip substrate used in the present work has an array of circular wells (diameter $\approx 70\unit{\mu m}$ and depth $\approx 1.4\unit{\mu m}$) that we microfabricated. To assemble the device, we first soaked the coverslip substrate in $1\unit{wt\%}$ solution of 3-(2-aminoethyl aminopropyl) trimethoxysilane (Shin-Etsu Chemical) and then treated it with biotin solution.
We also prepared a bilayer membrane of EMPS, which comprises a biotin-coated polyethylene terephthalate porous membrane (taken from Transwell 3450, Corning, with a nominal pore size of $0.4\unit{\mu m}$) and a streptavidin-decorated cellulose membrane (Spectra/Por 7, Repligen, Waltham, MA, molecular weight cut-off 25,000). Before the experiment, we put $1\unit{\mu L}$ of the bacterial suspension ($\mathrm{OD}_{600}=0.1$, prepared by the aforementioned method) on top of wells on the substrate and attached the bilayer membrane to confine bacteria in the wells below the membrane, by biotin-streptavidin bonding between the cellulose membrane and the coverslip.
Then we placed a double-sided tape (also acting as a spacer) of thickness $100\unit{\mu m}$ on the coverslip, enclosing the micropatterned region, and attached a PDMS pad on the double-sided tape.
This completes the fabrication of EMPS.

\subsection*{Observation of bacteria}
The assembled EMPS device was placed on the stage of an inverted optical microscope (Leica DMi8), equipped with a 63x (NA=1.30) oil immersion objective and operated by the software Leica LasX.
During the experiment, bacteria in the microfluidic wells were kept supplied with TB medium containing $0.02\unit{wt\%}$ of surfactant Tween-20, through the bilayer porous membrane of EMPS.
This medium was infused from a syringe by a pump (NE-1000, New Era Pump Systems), at the flow rate of $60\unit{mL/h}$ for the first $5\unit{min}$ and $2\unit{mL/h}$ for the rest.
As a result, bacteria grew and proliferated throughout each experiment.
We monitored bacterial population in a well (diameter $71.2(5)\unit{\mu m}$) by phase-contrast microscopy, capturing images (pixel size $0.1724\unit{\mu m}$, optical resolution $0.258\unit{\mu m}$) by a charge-coupled device camera (DFC3000G, Leica).
For the main experiment presented in this work, there were initially three cells in the observed well, but after roughly five hours, the number increased to $\approx 800$.
We then started to repeat series of acquisition of 1000 images at an interval of $\approx 0.0263\unit{s}$, preceded by automatic focus adjustment (Leica adaptive focus control, single-shot mode).
We further grouped the images into sets of 500 consecutive images and statistical analysis was carried out for each group. 
Bacteria continued growing and proliferating during this set of image acquisitions too, leading to an increase of the area fraction $\phi$ (\supfigref{S-fig:density}).
Note that, because the timing of the transition varied a little among different samples, we used a single biological replicate for the main analysis presented in this paper.
For the uninterrupted observation shown in \vidref{1}, phase-contrast images were acquired at a regular interval of $1\unit{s}$ with automatic focus adjustment (continuous mode).

\subsection*{Image pre-processing}
Pre-processing of the phase-contrast images consists of the following two parts.
We corrected the effect of non-uniform illumination, by normalizing the image intensity with the local threshold intensity evaluated by \texttt{adaptthresh} function of MATLAB.
Also, unless otherwise stipulated, we cropped the images to the region of interest of size $259 \times 214$ pixels ($44.7\unit{\mu m} \times 36.9\unit{\mu m}$) near the center of the well, to avoid influence from the well perimeter.

\subsection*{Estimation of the area fraction} 
The area fraction $\phi$, i.e., the ratio of the area occupied by bacteria to the total area of the region of interest, was evaluated by binarizing the pre-processed phase-contrast images as follows.
The binarization threshold was set by a method called the 3-class fuzzy c-means clustering \cite{FuzzyBook}.
Then we evaluated the fraction of the darker pixels in the binarized images.
Using all 500 images of each data set, we evaluated the mean and the standard deviation of the fraction of the darker pixels, and used them as the most probable value and the error, respectively, of the area fraction $\phi$.

\subsection*{Static structure factor} 
The static structure factor, $S(\vec{q})$ is defined as the squared modulus of the Fourier transform of the image intensity $I(\Vec{r},t)$, i.e., $S(\vec{q}) = \langle |\text{FT}[I(\vec{r},t)|^2]\rangle_{t}$, where $\text{FT}[\cdots]$ is the 2D Fourier transform and $\vec{q} = (q_x, q_y)$ is the wavenumber vector. 

\subsection*{Orientation field} 
The coarse-grained orientation field of bacteria, $\theta(\vec{r},t)$ with $\vec{r}=(x,y)$, was obtained by the structure tensor analysis \cite{Jahne-book}.
From the intensity field $I(x,y)$ of a pre-processed phase-contrast image (here $t$ is omitted from the argument for simplicity), the structure tensor $J(x,y)$ is defined by
\begin{equation}
  J(x,y) = 
    \begin{bmatrix}
    \langle{(\Delta_x^\sigma I)^2}\rangle_\sigma & 
    \langle{(\Delta_x^\sigma I)(\Delta_y^\sigma I)}\rangle_\sigma\\
    \langle{(\Delta_y^\sigma I)(\Delta_x^\sigma I)}\rangle_\sigma &  
    \langle{(\Delta_y^\sigma I)^2}\rangle_\sigma
\end{bmatrix},
\end{equation}
where 
\begin{align}
 \langle g(x,y) \rangle_\sigma &= \sum_{x',y'} g(x',y') f_\sigma(x-x', y-y') \\
 \Delta_x^\sigma I &= \sum_{x',y'} g(x',y') \prt{f_\sigma}{x}(x-x', y-y') \\
 \Delta_y^\sigma I &= \sum_{x',y'} g(x',y') \prt{f_\sigma}{y}(x-x', y-y')
\end{align}
with a Gaussian kernel $f_\sigma(x, y) = \frac{1}{2\pi\sigma^2} e^{-\frac{x^2 + y^2}{2\sigma^2}}$ and $\sigma = 6\unit{pixels} = 1.03\unit{\mu m}$.
Then, the orientation field $\theta(x,y,t)$ was obtained by
\begin{equation}
 \theta(x,y,t) = \frac{1}{2} \tan^{-1} \left(\frac{2\langle{(\Delta_x^\sigma I)(\Delta_y^\sigma I)}\rangle_\sigma}{\langle{(\Delta_x^\sigma I)^2}\rangle_\sigma  - \langle{(\Delta_y^\sigma I)^2}\rangle_\sigma} \right). 
\end{equation}

\subsection*{Tracking bacteria}
To track the motion of individual bacteria, we first need to carry out cell segmentation of phase-contrast images. This was done by applying MiSiC \cite{Panigrahi.etal-e2021}, a deep learning-based method for the segmentation of bacteria, iteratively to each pre-processed phase-contrast image. More specifically, after each application of MiSiC, we removed the regions where cells were detected, added small noise to the image intensity, varied parameters that set the criteria of cell detection slightly and randomly, and applied MiSiC again, unless cells were already detected in a sufficient fraction of the area or MiSiC was already applied sufficiently many times (specifically, 300 times).
After this iteration, we manually corrected the segmentation result if necessary, at least for the cell we focused on and its neighbors for the results presented in \figref{fig3} and \vidref{6}, and for all cells in the field of view, except those on the boundary, for the results in \figref{fig5}.
Then, using the segmentation results of all images in a given data set, we applied Blair and Dufresne's particle tracking code available at \url{https://site.physics.georgetown.edu/matlab/}, which is based on the algorithm developed by Crocker, Grier, and Weeks \cite{Crocker.Grier-JCIS1996}.
For the results on the active fluid phase shown in \figref{fig5}, manual correction of the tracking result was also necessary for bacteria that happened to move over a long distance through the neighbors.
Note that the analysis shown in \supfigref{S-fig:cage2} and \vidref{7} did not use the method described here, but is a result of fully manual tracking.

\subsection*{Nonlinear fit}
All nonlinear fits presented in this work were carried out by using the Levenberg–Marquardt algorithm, known to solve non-linear least-squares problems reliably.
The uncertainties of the fitting results indicate the 95\% confidence interval.

\section{Appendix B: Mode-coupling theory exponent $\gamma$} \label{sec:MCT}

Our bacterial glass transitions show a rapid increase of relaxation times, $\tau_Q$ and $\tau_\theta$ for translational and orientational degrees of freedom, respectively, as the area fraction $\phi$ is increased (\figref[c]{fig2} symbols).
The observed dependence on $\phi$ is consistent with the power-law divergence that mode-coupling theories (MCT) predict \cite{Gotze-book,berthier2011theoretical,Reichman.Charbonneau-JSM2005}  (\figref[c]{fig2} dashed lines), 
\begin{equation}
\tau_Q \sim (\phi_\mathrm{c}^Q - \phi)^{-\gamma_Q}, \qquad \tau_\theta \sim (\phi_\mathrm{c}^\theta - \phi)^{-\gamma_\theta},  \label{eq:MCT}
\end{equation}
as well as with the Vogel-Fulcher-Tamman (VFT) law \cite{Gotze-book,berthier2011theoretical,hunter2012physics} (\supfigref{S-fig:VFT}).

Here we focus on the exponent $\gamma$ of the MCT power law, \eqref{eq:MCT}.
Our fitting gives $\gamma_Q = 1.6(3)$ and $\gamma_\theta=1.5(13)$ (see main text).
In contrast, for thermal systems, it is actually known that MCT generally gives $\gamma \geq \gamma_0 \equiv 1.76\dots$ \cite{Gotze-book}.
More precisely, for spherical particle systems near equilibrium, one may indeed prove $\gamma \geq \gamma_0$ \cite{Gotze-book}.
The situation is more involved for the aspherical case, where the same inequality has not been proven from first principles, but by using a diagonalization approximation of the MCT memory kernel, one can still show $\gamma \geq \gamma_0$ \cite{Schilling.Scheidsteger-PRE1997}.
This inequality has also been confirmed by simulations of different aspherical particle systems (e.g., \cite{Moreno.etal-JCP2005,Pfleiderer.etal-EL2008}), without any exception so far, to our knowledge.
Therefore, it is reasonable to consider that $\gamma \geq \gamma_0$ generally holds for systems near equilibrium.
Our estimate $\gamma_Q = 1.6(3)$ seems to violate this bound, thereby indicating the non-equilibrium nature of our bacterial system.
This observation raises some interesting questions that may deserve further investigation, such as clarifying conditions to violate the inequality $\gamma \geq \gamma_0$, how generally this violation takes place in non-equilibrium or active systems, etc.

\end{document}

\newbibstartnumber{44}

%\title{Supplementary Information for ``Emergence of bacterial glass: two-step glass transition in 2D bacterial suspension''}
\title{Supplementary Information for ``Emergence of bacterial glass''}

\author{Hisay Lama}
%\affiliation{Department of Physics,\! The University of Tokyo,\! 7-3-1 Hongo,\! Bunkyo-ku,\! Tokyo 113-0033,\! Japan}%
\affiliation{Department of Physics, The University of Tokyo, Tokyo, Japan}%

\author{Masahiro J. Yamamoto}
\affiliation{National Metrology Institute of Japan, National Institute of Advanced Industrial Science and Technology, Tsukuba, Japan}
\affiliation{Department of Physics, The University of Tokyo, Tokyo, Japan}%

\author{Yujiro Furuta}
%\affiliation{Department of Physics,\! Tokyo Metropolitan University,\! 1-1 Minami-Oosawa,\! Hachioji,\! Tokyo 192-0397,\! Japan}
\affiliation{Department of Physics, Tokyo Metropolitan University, Tokyo, Japan}%
%\affiliation{Department of Physics,\! Tokyo Institute of Technology,\! 2-12-1 Ookayama,\! Meguro-ku,\! Tokyo 152-8551,\! Japan}
\affiliation{Department of Physics, Tokyo Institute of Technology, Tokyo, Japan}%

\author{Takuro Shimaya}
\affiliation{Department of Physics, The University of Tokyo, Tokyo, Japan}%
\affiliation{Department of Physics, Tokyo Institute of Technology, Tokyo, Japan}%

\author{Kazumasa A. Takeuchi}
\email{kat@kaztake.org}
\affiliation{Department of Physics, The University of Tokyo, Tokyo, Japan}%
\affiliation{Department of Physics, Tokyo Institute of Technology, Tokyo, Japan}%

\date{\today}

\maketitle

\section*{Supplementary Figures}

\begin{figure}[h]
    \centering
    \includegraphics[width=\hsize]{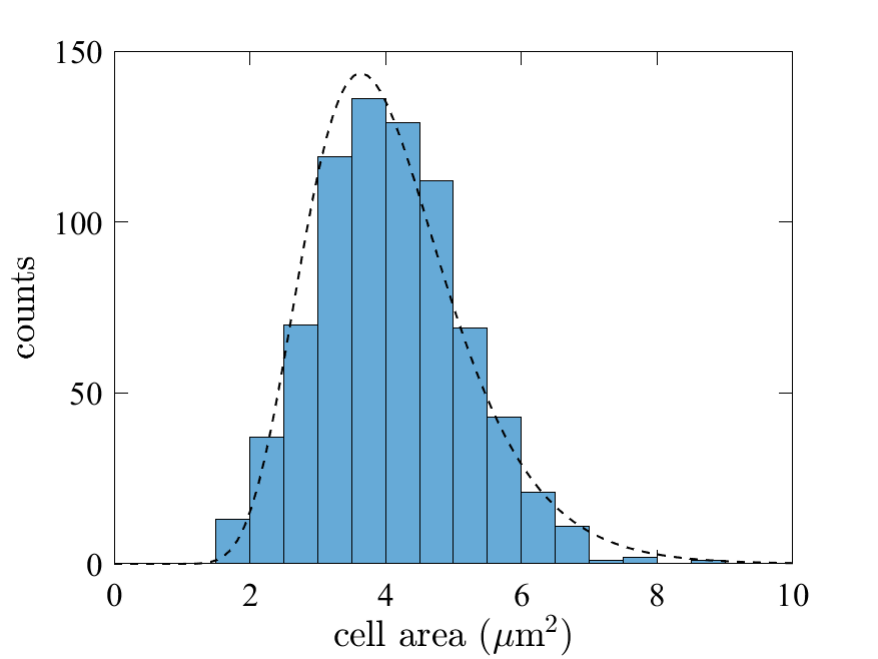}
    \caption{
    Histogram of cell areas. The cell areas were evaluated for $\phi = 0.784(7)$, from the first three frames where the cell segmentation was carried out. The dashed line shows the fitted log-normal distribution, whose probability density is given by $\frac{1}{\sqrt{2\pi}\sigma x}\exp\[-\frac{(\log x-\mu)^2}{2\sigma^2}\]$ with $\mu = 1.37(2)$ and $\sigma=0.281(15)$ (the value of $x$ is in the unit of $\mathrm{\mu m^2}$). Here the uncertainty corresponds to the 95\% confidence interval from the fit. The mean cell area is $4.08\unit{\mu m}$ and the polydispersity index is $1.07$. Note that a few filamentous cells and cells at the boundary of the region of interest were excluded from this histogram.
    }
    \label{fig:polydispersity}
\end{figure}

\begin{figure}[h]
    \centering
    \includegraphics[width=\hsize]{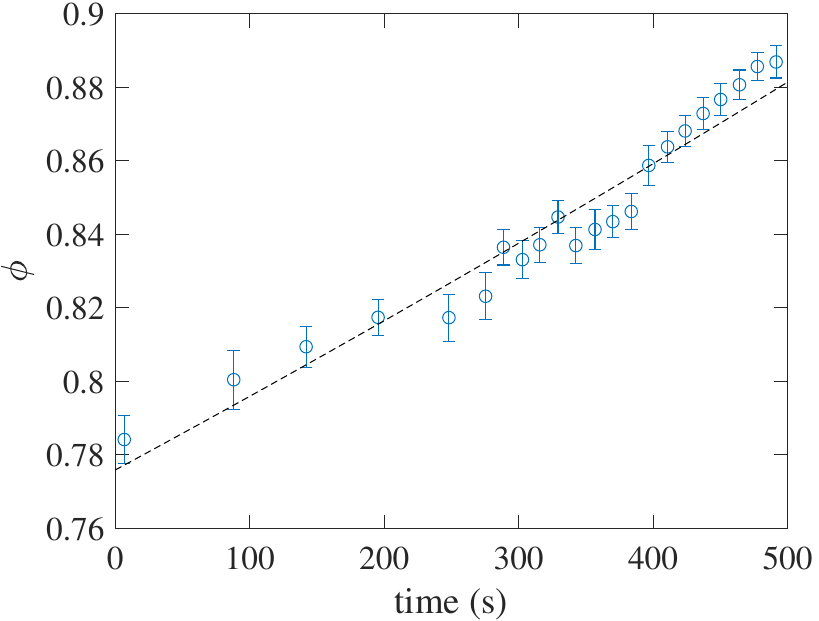}
    \caption{
    Growth of the area fraction $\phi$. The area fraction $\phi$ was evaluated for each group of 500 images recorded over $\approx 13\unit{s}$. The time point at the center is used in this plot. The dashed line shows a fit to the exponential growth curve, $\phi(t) \propto 2^{t/T}$, which estimated the doubling time at $T = 45(5)\unit{min}$. Here the uncertainty corresponds to the 95\% confidence interval from the fit. Note that the origin of time is set to be the moment at which we started the acquisition of the first set of images, which was already roughly five hours after the start of the experiment. 
    }
    \label{fig:density}
\end{figure}

\begin{figure*}[h]
    \centering
    \includegraphics[width=\hsize]{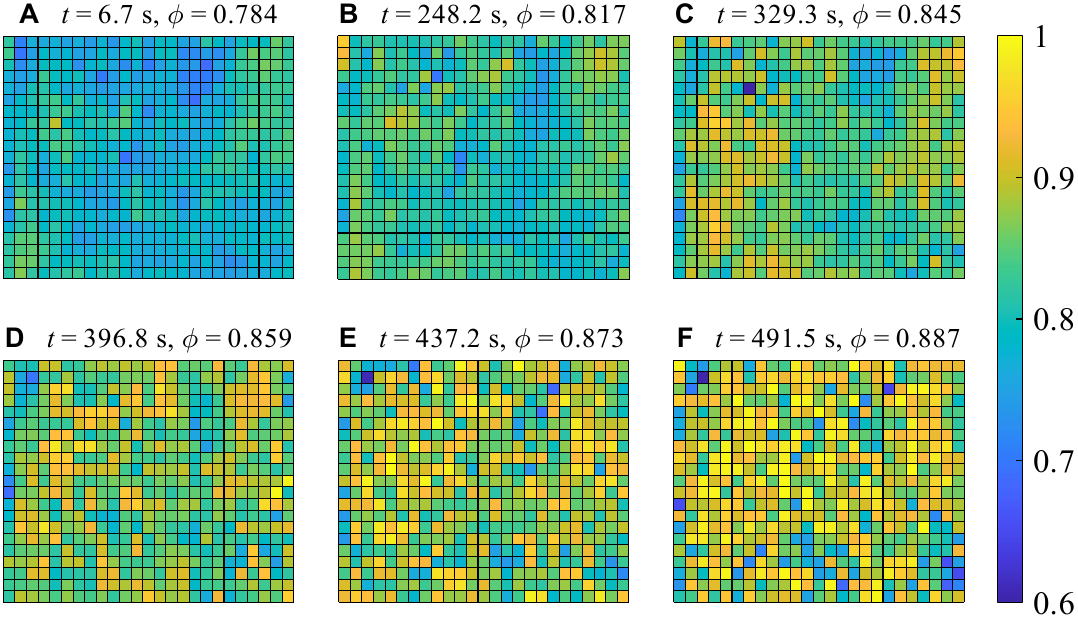}
    \caption{
    Spatial dependence of the area fraction $\phi$. The area fraction $\phi$ was evaluated for each group of 500 images recorded over $\approx 13\unit{s}$, locally in a mesh composed of regions of $10 \times 10$ pixels ($1.724\unit{\mu m} \times 1.724\unit{\mu m}$). Above each panel, the time at the center of each time interval and the mean value of $\phi$ (averaged over the time interval and the region of interest) are displayed. Note that the origin of time is set to be the moment at which we started the acquisition of the first set of images, which was already roughly five hours after the start of the experiment. These indicate that uniform growth was indeed realized in our device.
    }
    \label{fig:SpatialPhi}
\end{figure*}

\begin{figure*}[p]
    \centering
    \includegraphics[width=\hsize]{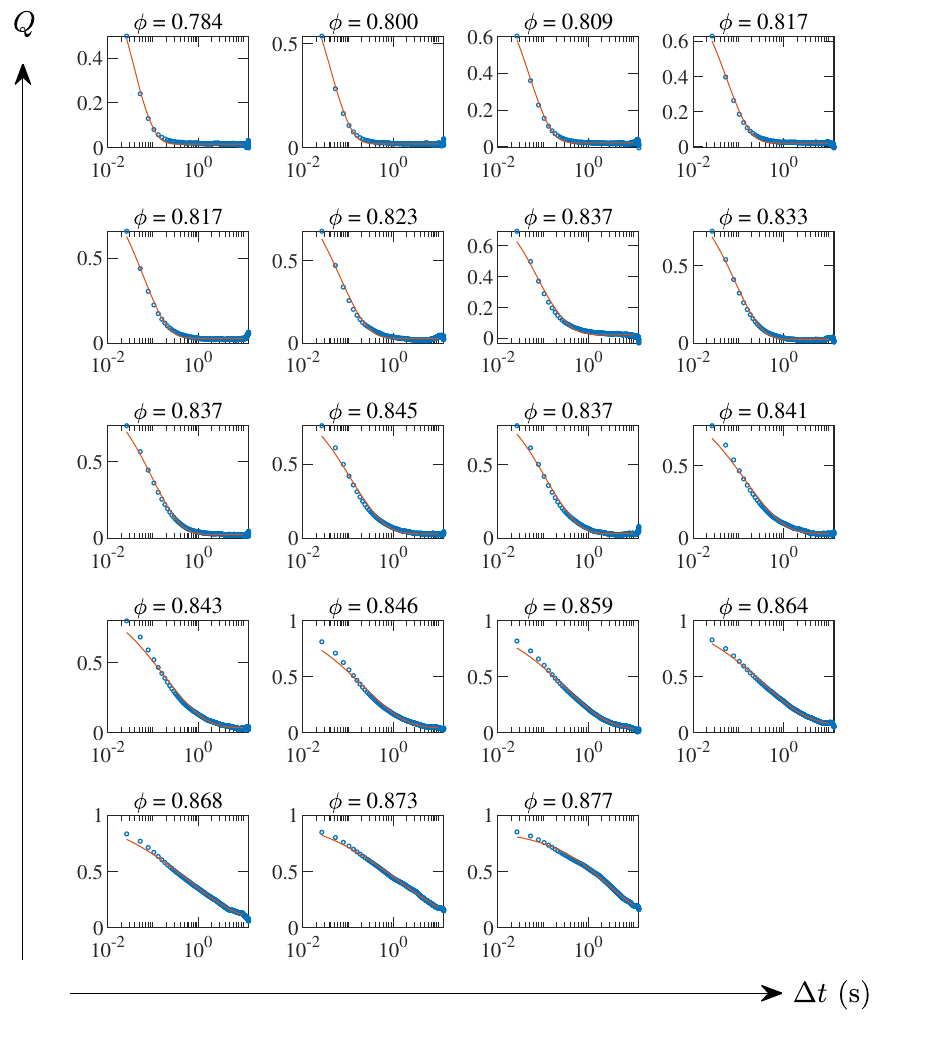}
    \caption{
    Results of fitting of the overlap function $Q(\Delta t)$ by a stretched exponential function.
    Each data was fitted by a stretched exponential function plus an offset, $Q(\Delta t) = f_Q e^{-(\Delta t/\tau_Q)^{\beta_Q}} + a_Q$.
    See \supfigref{fig:Beta} for the obtained values of $\beta_Q$.
    }
    \label{fig:StretchedExpQ}
\end{figure*}

\begin{figure}[h]
    \centering
    \includegraphics[width=\hsize]{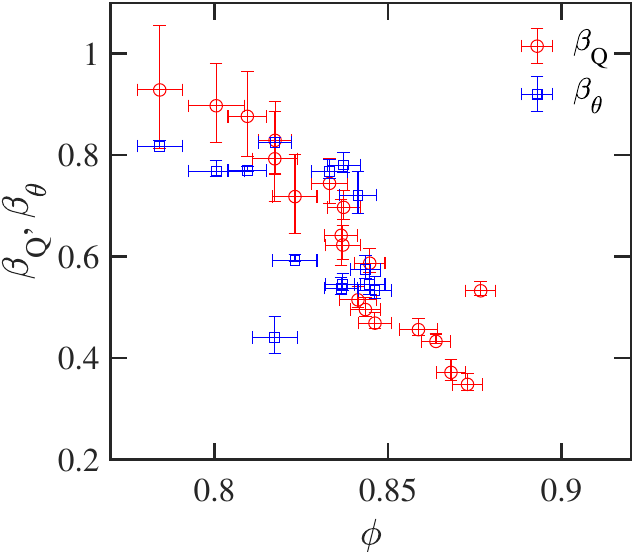}
    \caption{
    Stretched exponential parameter $\beta$.
    The estimates of $\beta_Q$ and $\beta_\theta$ from the stretched exponential fitting, $Q(\Delta t) = f_Q e^{-(\Delta t/\tau_Q)^{\beta_Q}} + a_Q$ and $C_\theta(\Delta t) = f_\theta e^{-(\Delta t/\tau_\theta)^{\beta_\theta}} + a_\theta$ (\supfigref{fig:StretchedExpQ} and \supfigref{fig:StretchedExpTheta}, respectively) are shown. The exponent values are closer to $1$ for lower area fractions $\phi$, indicating that the relaxation becomes closer to that of simple liquids.
    }
    \label{fig:Beta}
\end{figure}

\begin{figure}[h]
    \centering
    \includegraphics[width=\hsize]{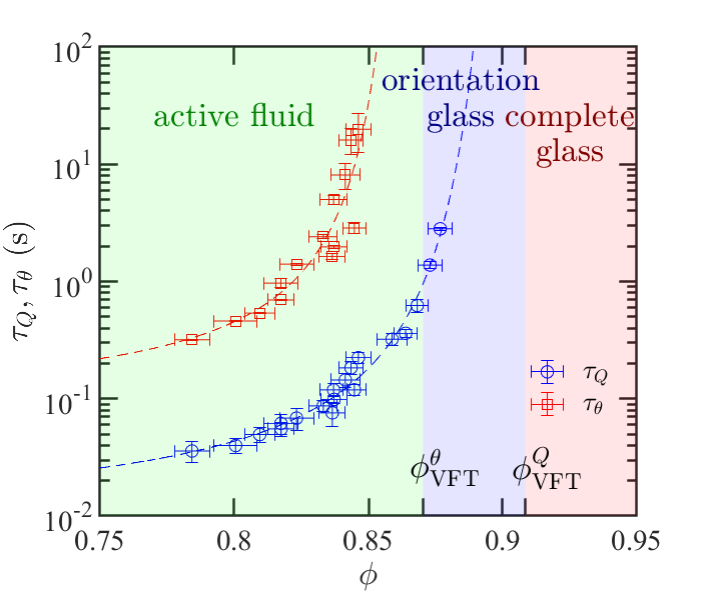}
    \caption{
    Vogel-Fulcher-Tamman (VFT) fitting of the relaxation times. The data of $\tau_Q$ and $\tau_\theta$ displayed in \figref[c]{M-fig2} are fitted here with the VFT law, $\tau \sim \exp\left(\frac{c\phi}{\phi_\mathrm{VFT}-\phi}\right)$ (dashed lines).
    The transition points are evaluated at $\phi_\mathrm{VFT}^Q = 0.908(12)$ for the translational relaxation and $\phi_\mathrm{VFT}^\theta = 0.870(35)$ for the orientational relaxation. The two-step transition scenario is also confirmed by the VFT fitting.
    }
    \label{fig:VFT}
\end{figure}

\begin{figure*}[p]
    \centering
    \includegraphics[width=\hsize]{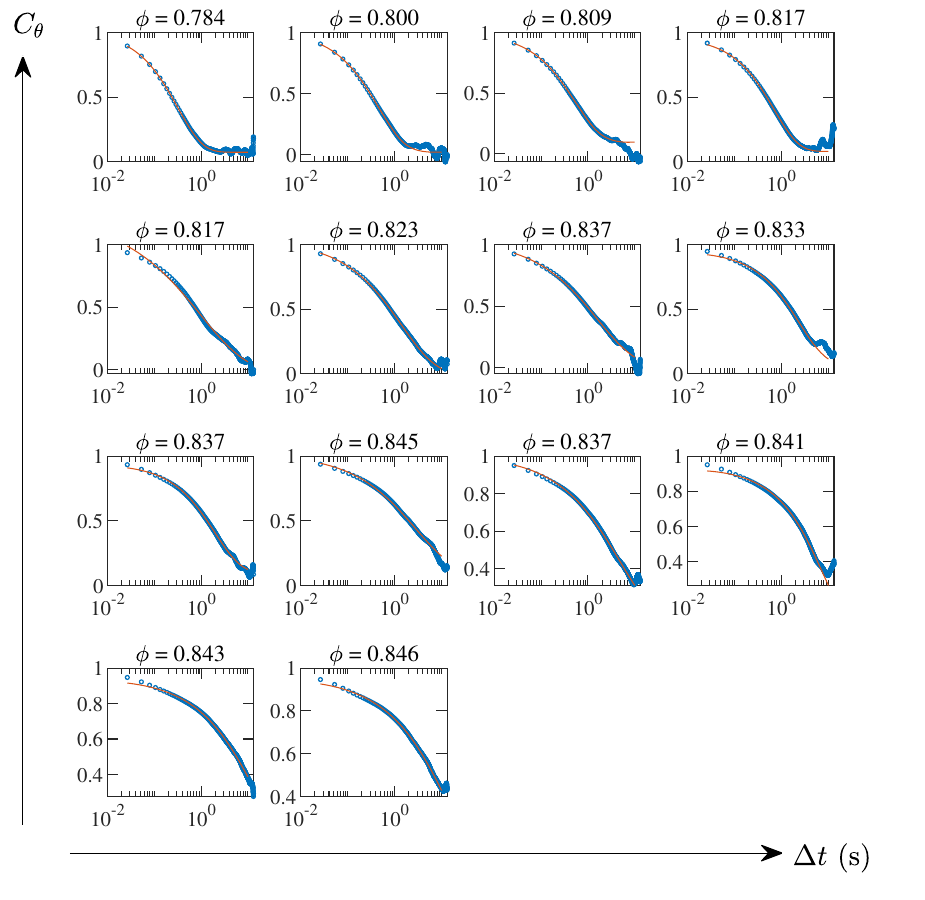}
    \caption{
    Results of fitting of the orientational correlation function $C_\theta(\Delta t)$ by a stretched exponential function.
    Each data was fitted by a stretched exponential function plus an offset, $C_\theta(\Delta t) = f_\theta e^{-(\Delta t/\tau_\theta)^{\beta_\theta}} + a_\theta$.
    See \supfigref{fig:Beta} for the obtained values of $\beta_\theta$.
    }
    \label{fig:StretchedExpTheta}
\end{figure*}

\begin{figure*}[hp]
    \centering
    \includegraphics[width=0.8\hsize]{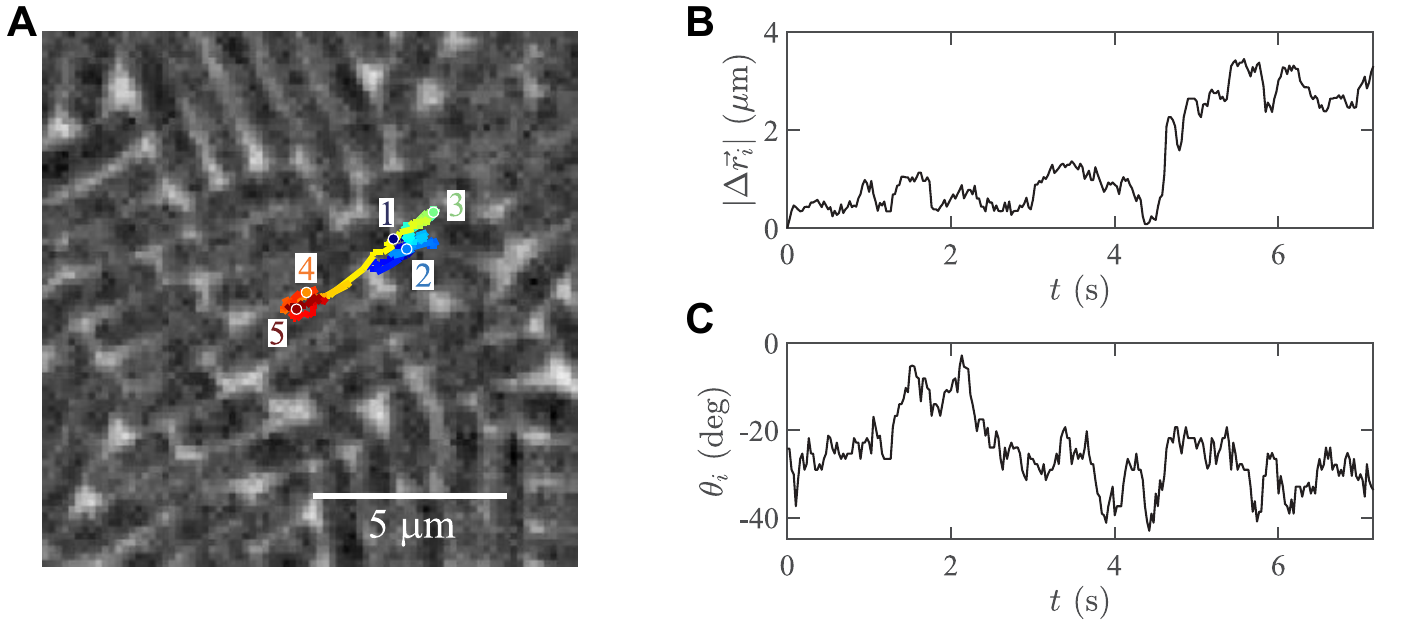}
    \caption{
    A cage escape event in the orientation glass phase, $\phi = 0.873(4)$, different from the one shown in \figref[a-c]{M-fig3}. See also \vidref{7}.
    In this event, the cell did not move across a border of microdomains, unlike the event shown in \figref[a-c]{M-fig3} and \vidref{6}.
    A) Trajectory of the single cell for $0 \leq t \leq 7.16\unit{s}$, drawn on the phase-contrast image taken at the last time frame. Note that the origin of time is different from that used in \figref[a-c]{M-fig3}. The positions at $t = 0, 1.76, 3.53, 5.29, 7.06\unit{s}$ are shown by colored disks with labels $1, 2, \cdots, 5$, respectively.
    B,C) Time series of the displacement from the initial position, $|\Delta\vec{r}_i(t)|$ (B), and that of the orientation $\theta_i(t)$ (C) of the cell tracked in panel A.
    These time series show a cage escape event during $4.5\unit{s} \lesssim t \lesssim 4.7\unit{s}$.
    }
    \label{fig:cage2}
\end{figure*}

\begin{figure*}[hp]
    \centering
    \includegraphics[width=0.8\hsize]{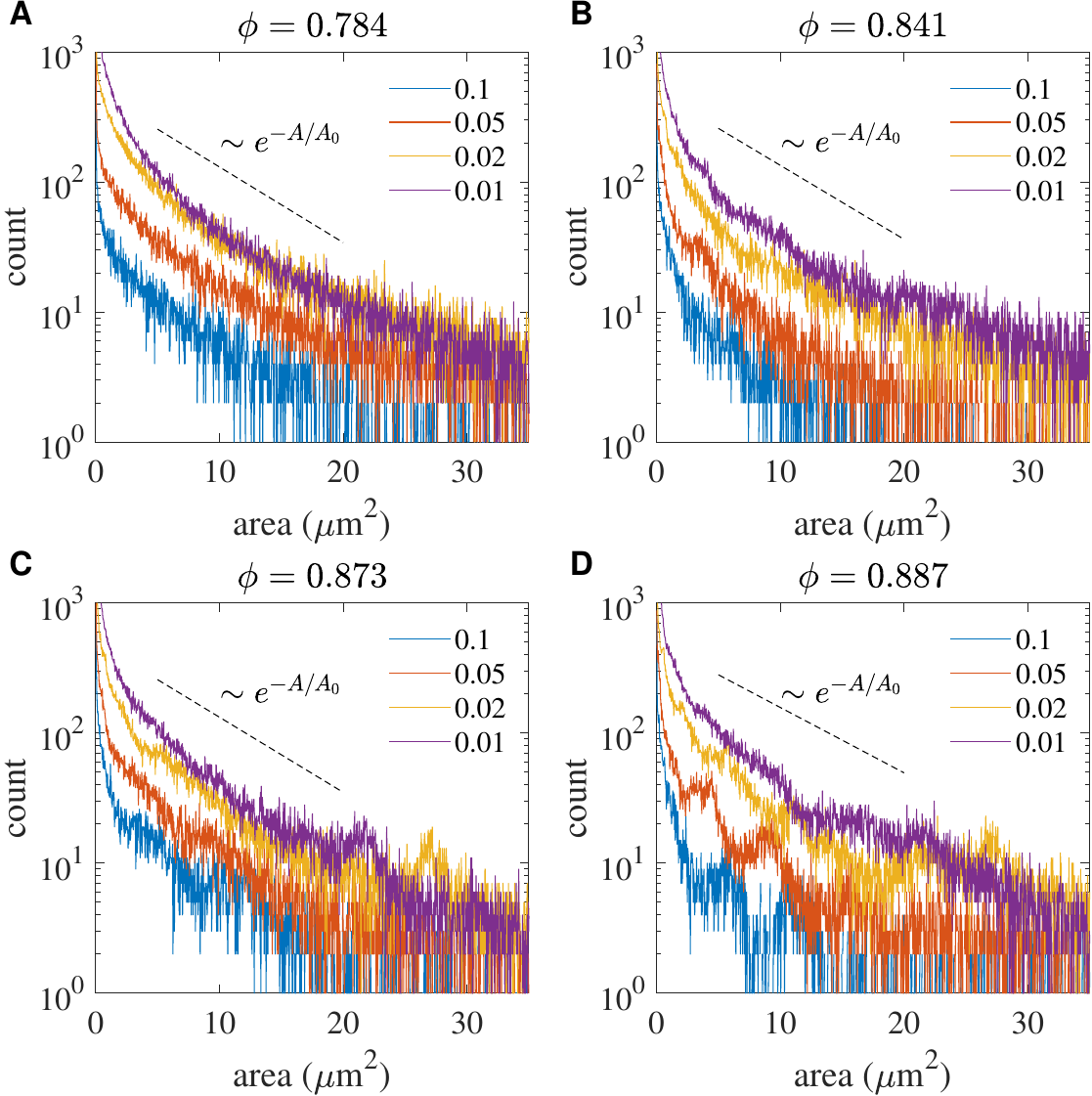}
    \caption{
    Microdomain size distributions for different thresholds of $|\nabla\theta(\vec{r},t)|^2$ (legend, in the unit of $\mathrm{(rad/\mu m)^2}$) and for different area fractions $\phi$.
    The dashed lines indicate the results of the fitting using all the displayed curves for each $\phi$.
    The resulting values of the characteristic area $A_0$ are $A_0 = 7.5 \pm 1.8\unit{\mu m^2}$ (\textbf{a}), $A_0 = 7.6 \pm 2.6\unit{\mu m^2}$ (\textbf{b}), $A_0 = 7.5 \pm 2.5\unit{\mu m^2}$ (\textbf{c}), and $A_0 = 8.6 \pm 4.8\unit{\mu m^2}$ (\textbf{d}).
    }
    \label{fig:DomainAreaDist}
\end{figure*}
\clearpage

\section*{Movie Captions}

\begin{description}
\item[Movie 1]
Uninterrupted Movie of bacteria undergoing glass transitions. The Movie starts from the active fluid phase where bacteria were actively swarming, and lasts until they become completely jammed. The Movie shows a central region of size $259 \times 214$ pixels ($44.7\unit{\mu m} \times 36.9\unit{\mu m}$) and played at 20 times the real speed. Scale bar: $10\unit{\mu m}$. Note that this Movie was taken from an experiment independent from the other observations, using a different substrate (well diameter $76.3(4)\unit{\mu m}$, depth $\approx 1.4\unit{\mu m}$).
\item[Movie 2]
Movie of bacteria at $\phi = 0.784(7)$ (active fluid phase). Scale bar $5\unit{\mu m}$. Played at real speed.
\item[Movie 3]
Movie of bacteria at $\phi = 0.841(5)$ (active fluid phase, close to $\phi_\mathrm{c}^\theta$). Scale bar $5\unit{\mu m}$. Played at real speed.
\item[Movie 4]
Movie of bacteria at $\phi = 0.873(4)$ (orientation glass). Scale bar $5\unit{\mu m}$. Played at real speed.
\item[Movie 5]
Movie of bacteria at $\phi = 0.887(4)$ (complete glass). Scale bar $5\unit{\mu m}$. Played at real speed.
\item[Movie 6]
The cage escape event in the orientation glass phase, $\phi = 0.873(4)$, shown in \figref[a-c]{M-fig3}.
The left panel shows the trajectory of the single cell drawn on the phase-contrast image. The right panels show the time series of the displacement from the initial position, $|\Delta\vec{r}_i(t)|$ (top) and that of the orientation $\theta_i(t)$ (bottom) of the cell shown in the left panel. The cage escape event took place during $5\unit{s} \lesssim t \lesssim 7\unit{s}$. 
\item[Movie 7]
Another cage escape event in the orientation glass phase, $\phi = 0.873(4)$, shown in \supfigref{fig:cage2}.
The left panel shows the trajectory of the single cell drawn on the phase-contrast image. The right panels show the time series of the displacement from the initial position, $|\Delta\vec{r}_i(t)|$ (top) and that of the orientation $\theta_i(t)$ (bottom) of the cell shown in the left panel. The cage escape event took place during $4.5\unit{s} \lesssim t \lesssim 4.7\unit{s}$.
\item[Movie 8]
Structure and evolution of microdomains at $\phi = 0.784(7)$ (active fluid phase). The left and right panels show the orientation field $\theta(\vec{r},t)$ and its gradient squared $|\nabla\theta(\vec{r},t)|^2$, respectively, overlaid on the phase-contrast image. Played at real speed. See also \figref{M-fig4}.
\item[Movie 9]
Structure and evolution of microdomains at $\phi = 0.841(5)$ (active fluid phase, close to $\phi_\mathrm{c}^\theta$). The left and right panels show the orientation field $\theta(\vec{r},t)$ and its gradient squared $|\nabla\theta(\vec{r},t)|^2$, respectively, overlaid on the phase-contrast image. Played at real speed. See also \figref{M-fig4}.
\item[Movie 10]
Structure and evolution of microdomains at $\phi = 0.873(4)$ (orientation glass). The left and right panels show the orientation field $\theta(\vec{r},t)$ and its gradient squared $|\nabla\theta(\vec{r},t)|^2$, respectively, overlaid on the phase-contrast image. Played at real speed. See also \figref{M-fig4}.
\item[Movie 11]
Structure and evolution of microdomains at $\phi = 0.887(4)$ (complete glass). The left and right panels show the orientation field $\theta(\vec{r},t)$ and its gradient squared $|\nabla\theta(\vec{r},t)|^2$, respectively, overlaid on the phase-contrast image. Played at real speed. See also \figref{M-fig4}.
\end{description}

%\begin{thebibliography}{99}
%\end{thebibliography}

%\bibliography{ref}